\documentclass{aa}  

\usepackage{graphicx}
\usepackage{natbib}
\usepackage{txfonts}
\usepackage{hyperref}
\usepackage{amssymb}
\usepackage{mathrsfs}

\begin{document} 


   \title{Photometric detection of internal gravity waves in upper main-sequence stars}
   \subtitle{IV. Comparable stochastic low-frequency variability in SMC, LMC, and \\ Galactic massive stars}

   \titlerunning{SLF variability in SMC and LMC massive stars}
   \authorrunning{D. M. Bowman et al.}
   
        \author{Dominic M. Bowman\inst{1,2} 
        \and
        Pieterjan Van Daele\inst{1,2}
        \and
        Mathias Michielsen\inst{2}
        \and
        Timothy Van Reeth\inst{2}
          }

         \institute{School of Mathematics, Statistics and Physics, Newcastle University, Newcastle upon Tyne, NE1 7RU, UK \\
              \email{dominic.bowman@newcastle.ac.uk}
         \and
         Institute of Astronomy, KU Leuven, Celestijnenlaan 200D, B-3001 Leuven, Belgium
         }

   \date{Received 8 July 2024; accepted 21 October 2024}


\abstract
   {Massive main-sequence stars have convective cores and radiative envelopes, but can also have sub-surface convection zones caused by partial ionisation zones. However, the convective properties of such regions strongly depend on opacity and therefore a star's metallicity. Non-rotating 1D evolution models of main-sequence stars between $7 \leq M \leq 40$~M$_{\odot}$ and the metallicity of the Small Magellanic Cloud (SMC) galaxy suggest tenuous (if any) sub-surface convection zones when using the Rayleigh number as a criterion for convection owing to their substantially lower metallicity compared to Galactic massive stars.}
   {We test whether massive stars of different metallicities both inside and outside of asteroseismically calibrated stability windows for sub-surface convection exhibit different properties in stochastic low-frequency (SLF) variability. Thus, we aim to constrain the metallicity dependence of the physical mechanism responsible for SLF variability commonly found in light curves of massive stars.}
   {We extracted customised light curves from the ongoing NASA Transiting Exoplanet Survey Satellite (TESS) mission for a sample of massive stars using an effective point spread function (ePSF) method, and compared their morphologies in terms of characteristic frequency, $\nu_{\rm char}$, and amplitude using a Gaussian process (GP) regression methodology.}
   {We demonstrate that the properties of SLF variability observed in time series photometry of massive stars are generally consistent across the metallicity range from the Milky Way down to the SMC galaxy, for stars both inside and outside of the sub-surface stability windows based on the Rayleigh number as a criterion for convection.}
   {We conclude that non-rotating 1D stellar structure models of sub-surface convection cannot alone be used to explain the mechanism giving rise to SLF variability in light curves of massive stars. Additionally, the similar properties of SLF variability across a wide range of metallicity values, which follow the same trends in mass and age in the Hertzsprung--Russell (HR) diagram at both high and low metallicity, support a transition in the dominant mechanism causing SLF variability from younger to more evolved stars. Specifically, core-excited internal gravity waves (IGWs) are favoured for younger stars lacking sub-surface convection zones, especially at low metallicity, and sub-surface convection zones are favoured for more evolved massive stars.}


   \maketitle


\section{Introduction}
\label{subsection: intro}

Our ability to ascertain precise and accurate constraints on the physical processes within massive stars, such as chemical mixing and angular momentum transport, has been revolutionised by studying stellar variability and applying forward asteroseismic modelling (see \citealt{Bowman2020c} for a review). Identifying coherent self-excited pulsation modes allow one to perform linear asteroseismology by quantitatively comparing observed pulsation mode frequencies to their theoretical counterparts calculated from a grid of stellar structure models \citep{ASTERO_Book, Aerts2021a}. Coherent pulsations are generally grouped into pressure (p) and gravity (g) modes based on their restoring forces that respectively probe the envelope and deep interior of a massive main-sequence star. Such coherent pulsations are excited by a periodic heat-engine mechanism operating within the iron-nickel opacity bump at 200\,000~K \citep{Dziembowski1993e, Dziembowski1993f, Miglio2007a, Szewczuk2017a}. Important constraints from forward asteroseismic modelling of coherent pulsations include interior mixing and rotation profiles, magnetic field strength and geometry, and the efficiency of angular momentum transport \citep{Aerts2003d, Dupret2004b, Pamyat2004, Dziembowski2008, Briquet2012, Salmon2022b, Lecoanet2022a, Burssens2023a}.

The necessary data for asteroseismic analyses of massive stars are long-duration time series, either photometric or spectroscopic, spanning at least a few months, but ideally several years \citep{Bowman2023b}. Given that high-cadence spectroscopic time series are quite expensive when assembled with ground-based telescopes (see e.g. \citealt{Aerts2003d}), asteroseismology using space-mission light curves has largely driven the field during the last two decades. Space missions with an asteroseismic component include CoRoT \citep{Auvergne2009}, BRITE \citep{Weiss2014, Weiss2021a}, \textit{Kepler}/K2 \citep{Borucki2010, Koch2010, Howell2014}, and the Transiting Exoplanet Survey Satellite (TESS; \citealt{Ricker2015}), which together have revealed a diversity of variability mechanisms at work in massive stars (see e.g. \citealt{Degroote2009b, Bowman2019b, Burssens2020a, Zwintz2024a}). 

Recently, a new form of asteroseismic signal called stochastic low-frequency (SLF) variability was discovered in many massive stars \citep{Bowman2019a, Bowman2019b}, and was shown to directly probe a massive star's mass and age \citep{Bowman2020b, Bowman2022b}. This SLF variability is seemingly ubiquitous in Galactic massive stars with light curve data of sufficiently high photometric precision. After an initial detection in three O stars observed by CoRoT \citep{Blomme2011b}, the first large-scale analysis of SLF variability among massive stars revealed a common morphology for dwarf stars with spectral types earlier than B9 \citep{Bowman2019a}. The sample was expanded to include hundreds of massive stars across the sky using photometry from the K2 mission in addition to dozens of blue supergiants in the Large Magellanic Cloud (LMC) galaxy for the first time \citep{Bowman2019b}. Later, combining TESS mission light curves and high-resolution spectroscopy, it was shown that a strong correlation exists between the properties of SLF variability and the location of stars in the Hertzsprung--Russell (HR) diagram \citep{Bowman2020b}. Recently, a novel method using Gaussian process (GP) regression confirmed this correlation, and revealed a transition from stochastic to quasi-periodic variability exists for massive Galactic stars during the main sequence \citep{Bowman2022b}. For stars without coherent heat-engine pulsation modes it is not possible to apply traditional forward asteroseismic modelling, thus SLF variability has motivated the advent of gravity wave asteroseismology \citep{Bowman2023b}.

Various theoretical and multi-dimensional hydrodynamical model predictions, which are not mutually exclusive, for explaining SLF variability in massive stars exist within the literature. Plausible mechanisms include internal gravity waves (IGWs) excited stochastically at the boundary of convective cores \citep{Rogers2013b, Rogers2015, Edelmann2019a, Horst2020a, Varghese2023a, Ratnasingam2019a, Ratnasingam2020a, Ratnasingam2023a, Vanon2023a, Herwig2023a, Thompson_W_2024a}, waves and/or turbulence caused by sub-surface convection zones \citep{Cantiello2009a, Cantiello2021b, Schultz2022a}, and inhomogeneities in the winds of massive stars \citep{Krticka2018e, Krticka2021b}. However, the last of these mechanisms is expected to play a negligible role for late O-type and early B-type main-sequence stars because of their optically thin and weak winds, especially at low metallicity. The explanation of SLF variability is hotly debated in the literature. Some hydrodynamical simulations suggest that IGWs excited by core convection reach the surface with smaller amplitudes than the observed amplitudes for SLF variability, which suggests that it arises from another convective region \citep{Lecoanet2019a, Lecoanet2021a, LeSaux2023a, Anders_E_2023b}. However, owing to the numerical challenges, hydrodynamical studies have generally not yet been performed using 3D spherical geometry with comparable rotations rates to observed stars. Moreover, some studies have shown that higher angular degree IGWs (i.e. $\ell \gtrsim 30$) make a significant contribution to mixing in the deep interiors of massive stars, and can potentially explain the shape of SLF variability at a star's surface \citep{Herwig2023a, Thompson_W_2024a}. This means that predictions from damping of high-$\ell$ in simulations can be quite sensitive to the numerical set-up, which in turn can make a quantitative comparison of different simulations to observations fairly challenging.

Recently, \citet{Jermyn2022a} demonstrated using non-rotating 1D stellar structure models that late O and early B main-sequence stars at low metallicity may not have sub-surface convection zones when the Rayleigh number is considered as a necessary criterion for convection. This calls into question whether sub-surface convection zones in the envelopes of massive stars at low metallicity are really regions with convection as the dominant transport mechanism and whether they can excite IGWs. Regardless, IGWs generated by turbulent core convection are not mutually exclusive with IGWs generated by sub-surface convection zones. Observations suggest a possible transition between IGWs excited by core convection being dominant at the zero-age main sequence (ZAMS) to IGWs excited by sub-surface convection being dominant at the terminal-age main sequence (TAMS) in the HR~diagram for Galactic massive stars \citep{Bowman2022b}. This transition remains unexplored outside of Galactic massive stars.

For this work, paper~IV in the current series, we used new TESS observations and advanced light curve extraction techniques applied to massive stars in the Small Magellanic Cloud (SMC) galaxy and compared them to stars in the LMC and the Milky Way. Our specific goal was to test whether a common morphology exists for SLF variability across a wide range of metallicities, keeping in mind the possible absence of sub-surface convection zones in low-metallicity massive stars \citep{Jermyn2022a}. In Sect.~\ref{section: method theory} we describe the calculation of 1D stellar structure and evolution models to define sub-surface convection stability windows, and in Sect.~\ref{section: method obs} we describe our target star selection criteria and TESS light curve extraction methodology. In Sect.~\ref{section: results} we present and discuss our analysis of the light curves and SLF variability, and we conclude in Sect.~\ref{section: conclusions}.


\section{Sub-surface convection stability windows}
\label{section: method theory}

        \subsection{Rayleigh number as a criterion for convection}
        \label{subsection: Rayleigh}

        Recently, \citet{Jermyn2022a} investigated the properties of sub-surface convection zones using the 1D stellar evolution code {\sc MESA} \citep{Paxton2011, Paxton2013, Paxton2015, Paxton2018, Paxton2019, Jermyn2023a}. They investigated whether the envelopes of massive stars satisfy the Rayleigh number, Ra, as a criterion for convection, which is defined as

        \begin{equation}
                {\rm Ra} \equiv \frac{g \left(\nabla_{\rm rad} - \nabla_{\rm ad}\right) \delta r^3}{\nu \alpha} \left(\frac{\delta r}{h}\right) \left( \frac{4-3\beta}{\beta} \right)
                \label{equation: Rayleigh}
        ,\end{equation}

        \noindent where $\nabla_{\rm rad}$ and $\nabla_{\rm ad}$ are the radiative and adiabatic temperature gradients, $g$ is the gravitational acceleration, $\delta r$ is the thickness of the Ledoux-unstable layer, $\nu$ is the kinematic viscosity, $\alpha$ is the thermal diffusivity, $h$ is the pressure scale height, and $\beta$ is the radiation parameter (cf. Appendices B and D of \citealt{Jermyn2022a} and references therein). The Rayleigh number indicates the importance of buoyancy when compared to diffusive processes, such that when ${\rm Ra} < {\rm Ra_{\rm crit}}$ buoyancy is insufficient to overcome diffusive processes and convection cannot occur. 

        \citet{Jermyn2022a} used the Rayleigh number to test whether sub-surface convection zones, associated with partial ionisation zones of metals (e.g. iron), helium, and hydrogen, are present for stars spanning $7 \leq M \leq 40$~M$_{\odot}$ and from the onset of core-hydrogen burning (i.e. ZAMS) until core-hydrogen exhaustion (i.e. TAMS). Using Eq.~(\ref{equation: Rayleigh}) as a condition for convective instability, \citet{Jermyn2022a} demonstrated that massive stars with Galactic metallicity always have a sub-surface convection zone within their envelopes during the main sequence. However, for lower metallicities, such as stars in the SMC and LMC galaxies, there exists a parameter space in the HR~diagram for which only the convective core satisfies the Rayleigh number, such that there are no sub-surface convection zones. \citet{Jermyn2022a} defined these parameter spaces as `stability windows' in the HR diagram in which the partial ionisation zones of metals, hydrogen, and helium are not convection zones. Therefore, given the significant differences in the convective properties of stellar envelopes between metal-rich and metal-poor stars, the properties of SLF variability observed in the light curves of such stars may be different if these zones are acting as the physical mechanism responsible for SLF variability.

        
        \subsection{Asteroseismic MESA models}
        \label{subsection: MESA}

        In this work, we expand upon the previous study of \citet{Jermyn2022a} by including convective boundary mixing (CBM) and envelope mixing calibrated by asteroseismology. This is important because interior mixing has a large impact on the width of the main sequence in the HR diagram and on the maximum radius and age of a massive star at the TAMS. A growing number of asteroseismic studies have demonstrated the importance of a non-negligible amount of CBM in massive main-sequence stars \citep{Dupret2004b, Pamyat2004, Dziembowski2008, Briquet2011, Burssens2023a}. In particular, a non-zero amount of CBM facilitates unprocessed hydrogen from the envelope to be transported into the nuclear-burning core, thus extending the main-sequence lifetime by at least 25\% and displacing the TAMS to cooler effective temperatures in the HR~diagram \citep{Bowman2020c}.
        
        \begin{figure}
        \centering
        \includegraphics[width=0.9\columnwidth]{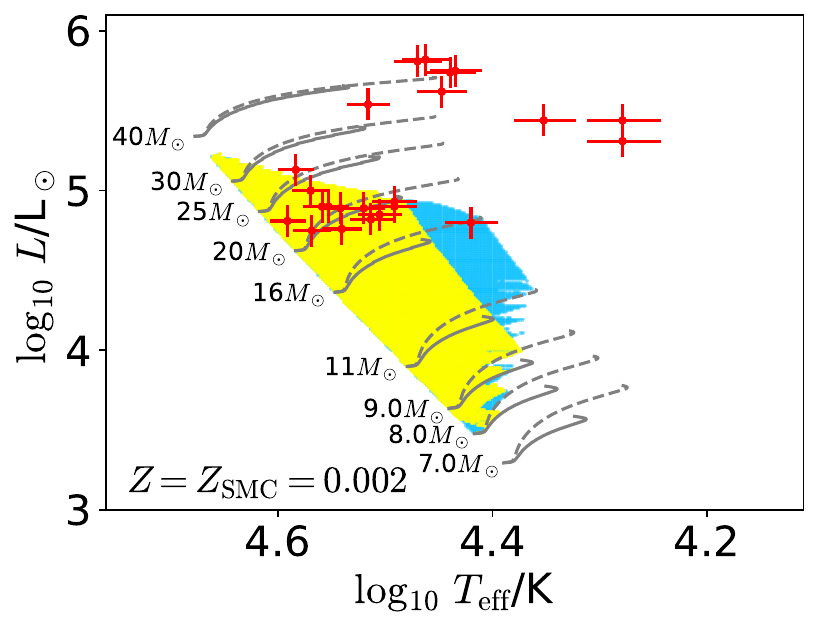}
        \includegraphics[width=0.9\columnwidth]{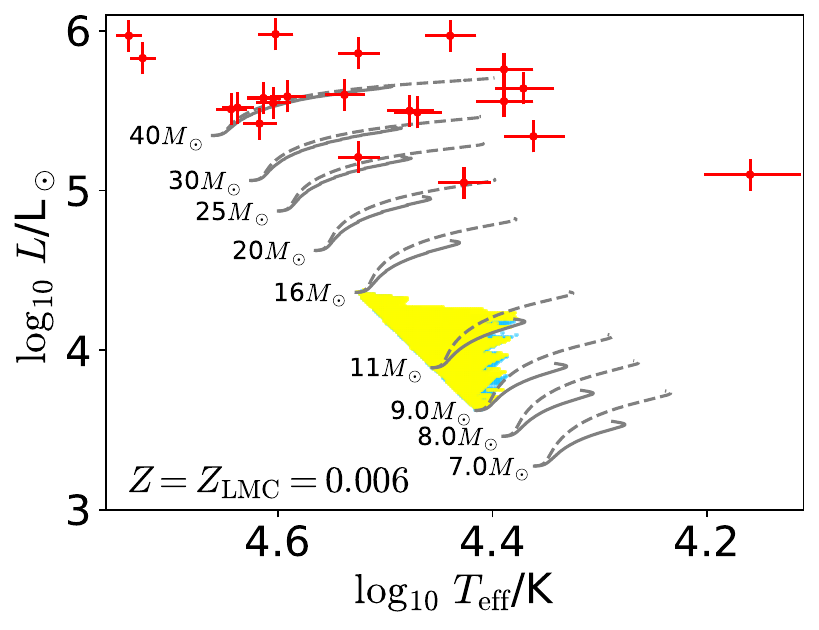}
        \caption{Hertzsprung--Russell (HR) diagram for SMC stars (i.e. $Z = 0.002$; top panel) and LMC stars (i.e. $Z = 0.006$; bottom panel) with evolutionary tracks of different masses between the ZAMS and the TAMS. The sub-surface convection zone stability windows as defined by the region for which ${\rm Ra} < 0.001$~Ra$_{\rm crit}$ (cf. Eq.~(\ref{equation: Rayleigh})) are shown in yellow for models with minimum mixing and blue for the additional models with maximum values of mixing. Evolutionary tracks with minimum and maximum mixing are shown as solid and dashed lines, respectively. The location of the sub-sample of stars with reliable TESS light curves, all of which have significant SLF variability, are shown as red points.}
        \label{figure: HRD}
        \end{figure}

        We calculated a grid of non-rotating evolutionary tracks using {\sc MESA} (v15140) for stars with masses $7 \leq M \leq 40$~M$_{\odot}$, a metal mass fraction consistent with the SMC (i.e. $Z = 0.002 \simeq 0.2$~Z$_{\odot}$), and ages that span from the ZAMS to the TAMS. We included a fixed amount of envelope mixing (i.e. $\log(D_{\rm env}) = 3.0$) and a range in CBM using the diffusive exponential mixing prescription (i.e. $0.0 \leq f_{\rm ov} \leq 0.3$) as inferred from detailed forward asteroseismic modelling of pulsating stars within and proximate to this parameter space (e.g. \citealt{Michielsen2021a, Pedersen2021a, Burssens2023a}). For completeness, we also calculated a second grid of {\sc MESA} models with the same range of parameters except that the metal mass fraction was consistent with the LMC (i.e. $Z = 0.006 \simeq 0.5$~Z$_{\odot}$).

        Our sub-surface convection zone stability windows that encompass the full range of minimum to maximum values of CBM inferred from asteroseismology are shown as the combined yellow and blue regions in Fig.~\ref{figure: HRD}. We chose to be conservative in the definition of the stability windows, and specifically show models that satisfy ${\rm Ra} < 0.001$~Ra$_{\rm crit}$ (cf. Eq.~(\ref{equation: Rayleigh})), thus demonstrating that the Rayleigh number criterion for the lack of convection is robustly satisfied. The transition just beyond the edge of the stability windows from ${\rm Ra} \sim$~Ra$_{\rm crit}$ to ${\rm Ra} < 0.001$~Ra$_{\rm crit}$ is extremely sharp in the HR~diagram. The stability windows cover a larger parameter space in the HR~diagram compared to those of \citet{Jermyn2022a} because of the non-zero amount of CBM. We show the evolutionary tracks with the minimum and maximum values of CBM as solid and dashed lines, respectively, which displaces the TAMS to cooler temperatures and higher luminosities because of the extended main-sequence lifetime being at least 25\% longer in massive stars that include non-zero CBM. As expected, the size of the stability window for SMC metallicity is much larger than for LMC metallicity \citep{Jermyn2022a}.


\section{TESS light curves of SMC and LMC massive stars}
\label{section: method obs}

        TESS is an all-sky time series photometric survey mission observing individual sectors of the sky for up to 27~d \citep{Ricker2015}. Its primary science case is to identify candidate exoplanet transits orbiting bright Galactic stars. TESS provides its entire field of view spanning four cameras as full-frame images (FFIs) at a cadence of 200~s starting from its second extended mission \citep{Jenkins2016b}. Previous FFI cadences were 30~min and 10~min in the nominal mission and first extended mission, respectively. Bias subtraction and flat fielding were performed before delivery of FFIs to the Mikulski Archive of Space Telescopes (MAST\footnote{\url{https://archive.stsci.edu/missions-and-data/tess}}). 

        \subsection{SMC and LMC sample selection}
        \label{subsection: sample}
        
\begin{table*}
\caption{Parameters of SMC massive stars proximate to and within the sub-surface convection zone stability window studied in this work.}
\begin{center}
\resizebox{0.99\textwidth}{!}{
\begin{tabular}{c c c c c c c c c}
\hline \hline
Simbad name     & TIC ID        & Gaia DR3      & $T_{\rm eff}$         & $\log_{10}(L / {\rm L}_{\odot})$        & $\nu_{\rm char}$      & PSD$_{\rm max}$                           & PSD$_{\rm min}$                               \\
                        &               &                       & (K)                   &                                                         & (d$^{-1})$            & ({\rm mmag$^2$}/{d$^{-1}$})             & ({\rm mmag$^2$}/{d$^{-1}$})           \\

\hline
\vspace{0.1cm}
AzV 16                                  &       180617038       &       4685854571724688768     &       $27200$ &       $5.75$  &       $0.33^{+0.08}_{-0.09}$  &       $2.43 \times 10^{-3}$ &       $3.29 \times 10^{-7}$   \\
\vspace{0.1cm}
AzV 18                                  &       180617057       &       4688950933906863488     &       $19000$ &       $5.44$  &       $0.35^{+0.17}_{-0.18}$  &       $1.06 \times 10^{-3}$ &       $2.57 \times 10^{-7}$   \\
\vspace{0.1cm}
AzV 83                                  &       181043369       &       4688966125248119808     &       $32800$ &       $5.54$  &       $0.88^{+0.13}_{-0.12}$  &       $7.69 \times 10^{-4}$ &       $4.74 \times 10^{-6}$   \\
\vspace{0.1cm}
Cl* NGC 346 MPG 12              &       181880133       &       4689016221756858624     &       $31000$ &       $4.93$  &       $0.96^{+0.20}_{-0.17}$  &       $4.26 \times 10^{-4}$ &       $1.10 \times 10^{-5}$   \\
\vspace{0.1cm}
AzV 202                                 &       181887485       &       4689019898248677888     &       $26300$ &       $4.80$  &       $2.04^{+1.97}_{-0.95}$  &       $1.68 \times 10^{-5}$ &       $4.25 \times 10^{-6}$   \\
\vspace{0.1cm}
Cl* NGC 346 ELS 46                      &       182294016       &       4690504582501410432     &       $39000$ &       $4.81$  &       $2.10^{+9.16}_{-1.74}$  &       $2.87 \times 10^{-3}$ &       $1.13 \times 10^{-4}$   \\
\vspace{0.1cm}
Cl* NGC 346 ELS 25                      &       182294030       &       4690504994818102016     &       $36200$ &       $4.90$  &       $0.73^{+0.15}_{-0.13}$  &       $3.02 \times 10^{-3}$ &       $3.64 \times 10^{-5}$   \\
\vspace{0.1cm}
AzV 267                                 &       182294155       &       4690507022042458624     &       $35700$ &       $4.90$  &       $1.47^{+6.93}_{-0.63}$  &       $5.30 \times 10^{-4}$ &       $3.26 \times 10^{-5}$   \\
\vspace{0.1cm}
AzV 264                                 &       182294249       &       4690520387983090432     &       $22500$ &       $5.44$  &       $0.40^{+0.08}_{-0.09}$  &       $1.17 \times 10^{-3}$ &       $3.59 \times 10^{-7}$   \\
\vspace{0.1cm}
AzV 251                                 &       182300503       &       4685990709381204864     &       $33100$ &       $4.89$  &       $2.69^{+10.02}_{-1.11}$ &       $9.92 \times 10^{-5}$ &       $6.16 \times 10^{-6}$   \\
\vspace{0.1cm}
Cl* NGC 346 MPG 682             &       182300967       &       4689015465842590464     &       $34800$ &       $4.89$  &       $0.73^{+3.54}_{-0.66}$  &       $4.71 \times 10^{-3}$ &       $2.99 \times 10^{-5}$   \\
\vspace{0.1cm}
AzV 372                                 &       182729695       &       4687411518857964032     &       $28000$ &       $5.62$  &       $0.48^{+0.09}_{-0.09}$  &       $5.46 \times 10^{-4}$ &       $2.50 \times 10^{-7}$   \\
\vspace{0.1cm}
AzV 429                                 &       182910549       &       4687532533854919680     &       $38300$ &       $5.13$  &       $1.03^{+0.24}_{-0.19}$  &       $1.39 \times 10^{-4}$ &       $6.14 \times 10^{-6}$   \\
\vspace{0.1cm}
AzV 468                                 &       183302393       &       4687248791131672960     &       $34700$ &       $4.76$  &       $7.78^{+16.26}_{-6.98}$ &       $4.66 \times 10^{-5}$ &       $1.07 \times 10^{-5}$   \\
\vspace{0.1cm}
AzV 479                                 &       183306464       &       4686408759936714752     &       $29000$ &       $5.82$  &       $1.05^{+0.18}_{-0.20}$  &       $2.36 \times 10^{-5}$ &       $1.46 \times 10^{-7}$   \\
\vspace{0.1cm}
AzV 472                                 &       183306676       &       4687241472507227520     &       $19000$ &       $5.31$  &       $0.49^{+0.08}_{-0.09}$  &       $2.54 \times 10^{-4}$ &       $1.87 \times 10^{-7}$   \\
\vspace{0.1cm}
AzV 488                                 &       183492668       &       4686413437158238208     &       $27500$ &       $5.74$  &       $0.56^{+0.10}_{-0.10}$  &       $1.75 \times 10^{-4}$ &       $7.58 \times 10^{-8}$   \\
\vspace{0.1cm}
{\rm [M2002]} SMC 81469         &       183979145       &       4686450580031889536     &       $31000$ &       $4.90$  &       $1.99^{+0.39}_{-0.30}$  &       $4.21 \times 10^{-5}$ &       $2.50 \times 10^{-6}$   \\
\vspace{0.1cm}
2dFS 3780                               &       183981879       &       4686448106130736256     &       $32000$ &       $4.85$  &       $1.39^{+0.36}_{-0.28}$  &       $1.58 \times 10^{-5}$ &       $2.39 \times 10^{-6}$   \\
\vspace{0.1cm}
2dFS 163                                        &       267429694       &       4688838203919272832     &       $32600$ &       $4.82$  &       $1.48^{+0.30}_{-0.25}$  &       $4.00 \times 10^{-4}$ &       $1.21 \times 10^{-5}$   \\
\vspace{0.1cm}
2dFS 3954                               &       303910981       &       4686266721070836608     &       $37000$ &       $4.75$  &       $2.95^{+16.79}_{-1.97}$ &       $4.29 \times 10^{-5}$ &       $1.12 \times 10^{-5}$   \\
\vspace{0.1cm}
AzV 461                                 &       402099597       &       4687470686324401536     &       $37100$ &       $5.00$  &       $0.36^{+0.12}_{-0.14}$  &       $5.76 \times 10^{-3}$ &       $4.50 \times 10^{-6}$   \\
\vspace{0.1cm}
AzV 456                                 &       402100664       &       4687251368112248960     &       $29500$ &       $5.81$  &       $0.83^{+0.21}_{-0.24}$  &       $2.95 \times 10^{-5}$ &       $2.04 \times 10^{-7}$   \\

\hline\hline
\end{tabular} }
\tablefoot{Spectroscopic parameters are from \citet{Vink2023a} and \citet{Hawcroft2024a}, and the characteristic frequency, $\nu_{\rm char}$, and its $2\sigma$ confidence interval determined using a Gaussian process (GP) regression methodology \citep{Bowman2022b}. The maximum and minimum values of the power spectral density (PSD) of the best-fit GP regression model, as proxies for $\alpha_0$ and $C_{\rm W}$, respectively, of the SLF variability are also provided. An average for the typical reported uncertainties are $\sigma\left(T_{\rm eff}\right) = 1500$~K and $\sigma\left(\log_{10}\left(L / {\rm L}_{\odot} \right) \right) = 0.1$~dex.}
\end{center}
\label{table: SMC stars}
\end{table*}

\begin{table*}
\caption{Parameters of LMC massive stars proximate to the sub-surface convection zone stability window studied in this work.}
\begin{center}
\resizebox{0.99\textwidth}{!}{
\begin{tabular}{c c c c c c c c c}
\hline \hline
Simbad name     & TIC ID        & Gaia DR3      & $T_{\rm eff}$         & $\log_{10}(L / {\rm L}_{\odot})$        & $\nu_{\rm char}$      & PSD$_{\rm max}$                           & PSD$_{\rm min}$                               \\
                        &               &                       & (K)                   &                                                         & (d$^{-1})$            & ({\rm mmag$^2$}/{d$^{-1}$})             & ({\rm mmag$^2$}/{d$^{-1}$})   \\

\hline

\vspace{0.1cm}
$[$L72$]$ LH 9$-$89     &       30275740                &       4662154018793644928     &       $26700$         &       $5.05$  &       $0.80^{+0.37}_{-0.31}$  &       $9.89 \times 10^{-3}$ &       $6.16 \times 10^{-6}$   \\
\vspace{0.1cm}
$[$ELS2006$]$ N11 051   &       30275973                &       4662156630134021376     &       $41400$         &       $5.42$  &       $2.50^{+1.11}_{-0.65}$  &       $3.67 \times 10^{-5}$ &       $1.64 \times 10^{-6}$   \\
\vspace{0.1cm}
Sk $-$66 18                     &       30312045                &       4662202878335589376     &       $40200$         &       $5.55$  &       $2.87^{+1.79}_{-0.82}$  &       $1.55 \times 10^{-5}$ &       $5.69 \times 10^{-7}$   \\
\vspace{0.1cm}
$[$ELS2006$]$ N11 046   &       30313090                &       4662144501145651328     &       $33500$         &       $5.21$  &       $1.17^{+1.10}_{-0.43}$  &       $4.43 \times 10^{-4}$ &       $5.24 \times 10^{-6}$   \\
\vspace{0.1cm}
Sk $-$68 41                     &       31105740                &       4661392533937464448     &       $24500$         &       $5.56$  &       $0.42^{+0.05}_{-0.06}$  &       $5.71 \times 10^{-4}$ &       $9.05 \times 10^{-8}$   \\
\vspace{0.1cm}
Sk $-$68 52                     &       31181554                &       4661289630893455488     &       $24500$         &       $5.76$  &       $0.32^{+0.06}_{-0.07}$  &       $1.18 \times 10^{-3}$ &       $8.90 \times 10^{-8}$   \\
\vspace{0.1cm}
Sk $-$65 22                     &       55758033                &       4662240708409788928     &       $33500$         &       $5.86$  &       $0.67^{+0.09}_{-0.10}$  &       $2.63 \times 10^{-4}$ &       $1.29 \times 10^{-7}$   \\
\vspace{0.1cm}
W61 28$-$5                      &       276936126       &       4657275137819689216     &       $44000$         &       $5.51$  &       $0.93^{+0.18}_{-0.15}$  &       $1.61 \times 10^{-4}$ &       $4.52 \times 10^{-6}$   \\
\vspace{0.1cm}
VFTS 72                         &       277300368       &       4657698454092124416     &       $54800$         &       $5.97$  &       $1.04^{+0.16}_{-0.16}$  &       $4.76 \times 10^{-5}$ &       $1.15 \times 10^{-6}$   \\
\vspace{0.1cm}
Sk~$-$66 17                     &       277103700       &       4660278492484698112     &       $29500$         &       $5.49$  &       $1.06^{+0.11}_{-0.10}$  &       $9.97 \times 10^{-5}$ &       $2.13 \times 10^{-7}$   \\
\vspace{0.1cm}
BI 237                          &       277105480       &       4660109236409051392     &       $53200$         &       $5.83$  &       $1.04^{+6.09}_{-0.29}$  &       $9.09 \times 10^{-6}$ &       $5.99 \times 10^{-7}$   \\
\vspace{0.1cm}
VFTS 244                                &       277300097       &       4657680552664841856     &       $41050$         &       $5.58$  &       $1.87^{+0.48}_{-0.39}$  &       $4.94 \times 10^{-4}$ &       $2.43 \times 10^{-5}$   \\
\vspace{0.1cm}
VFTS 355                                &       277300563       &       4657688695923883648     &       $43400$         &       $5.52$  &       $1.08^{+0.13}_{-0.11}$  &       $3.99 \times 10^{-5}$ &       $2.33 \times 10^{-6}$   \\
\vspace{0.1cm}
Sk $-$68 135                    &       277300709       &       4657700790554314752     &       $27500$         &       $5.97$  &       $0.24^{+0.09}_{-0.12}$  &       $2.99 \times 10^{-3}$ &       $6.42 \times 10^{-8}$   \\
\vspace{0.1cm}
Sk $-$71 35                     &       287436747       &       4651835961162249984     &       $23000$         &       $5.34$  &       $0.35^{+0.08}_{-0.10}$  &       $5.85 \times 10^{-3}$ &       $4.76 \times 10^{-7}$   \\
\vspace{0.1cm}
Sk $-$66 100                    &       373849172       &       4660223379469868416     &       $39000$         &       $5.59$  &       $2.75^{+4.79}_{-1.03}$  &       $1.60 \times 10^{-5}$ &       $3.99 \times 10^{-7}$   \\
\vspace{0.1cm}
BI 173                          &       373919576       &       4658104341338342528     &       $34500$         &       $5.60$  &       $2.42^{+4.29}_{-0.95}$  &       $1.34 \times 10^{-5}$ &       $3.21 \times 10^{-7}$   \\
\vspace{0.1cm}
RMC 109                         &       391746326       &       4660181151347435008     &       $14450$         &       $5.10$  &       $0.27^{+0.08}_{-0.10}$  &       $8.41 \times 10^{-4}$ &       $1.47 \times 10^{-7}$   \\
\vspace{0.1cm}
Sk $-$71 41                     &       391887962       &       4651834994760055040     &       $30000$         &       $5.50$  &       $0.77^{+0.09}_{-0.09}$  &       $1.43 \times 10^{-4}$ &       $2.55 \times 10^{-7}$   \\
\vspace{0.1cm}
Sk $-$68 140                    &       404768212       &       4657688313632701568     &       $23500$         &       $5.64$  &       $0.48^{+0.09}_{-0.10}$  &       $8.72 \times 10^{-4}$ &       $2.78 \times 10^{-7}$   \\
\vspace{0.1cm}
Sk $-$67 166                    &       425083410       &       4660121743354291328     &       $40000$         &       $5.98$  &       $1.39^{+0.11}_{-0.12}$  &       $4.33 \times 10^{-5}$ &       $1.33 \times 10^{-7}$   \\

\hline\hline
\end{tabular} }
\tablefoot{Spectroscopic parameters are from \citet{Vink2023a} and \citet{Hawcroft2024a}, and the characteristic frequency, $\nu_{\rm char}$, and its $2\sigma$ confidence interval determined using a Gaussian process (GP) regression methodology \citep{Bowman2022b}. The maximum and minimum values of the power spectral density (PSD) of the best-fit GP regression model, as proxies for $\alpha_0$ and $C_{\rm W}$, respectively, of the SLF variability are also provided. An average for the typical reported uncertainties are $\sigma\left(T_{\rm eff}\right) = 1500$~K and $\sigma\left(\log_{10}\left(L / {\rm L}_{\odot} \right) \right) = 0.1$~dex.}
\end{center}
\label{table: LMC stars}
\end{table*}

        As our goal is to investigate the occurrence and properties of SLF variability of massive stars at low metallicity, we require a bona fide sample of massive stars in the SMC and LMC galaxies to place in the HR~diagram. Massive stars, including dwarfs, giants, and supergiants of spectral types earlier than about B9 in the LMC have been studied spectroscopically thanks to Very Large Telescope (VLT) programmes with European Southern Observatory (ESO) instruments, such as VLT-Flames (PI: Smart) and its successor the VLT-Flames Tarantula Survey (VFTS; PI: Evans), and more recently the ongoing Binarity at Low Metallicity (BLOeM) survey for the SMC (PI: Shenar; \citealt{Shenar2024a}). Breakthroughs in studying LMC massive stars include the metallicity dependence of winds and mass loss \citep{Mokiem2007a}, empirical rotation rates \citep{Ramirez-Agudelo2013}, surface-nitrogen abundances and their relation to interior mixing \citep{Hunter_I_2008b}, and constraints on binary statistics \citep{Sana2012b}.
        
        The \textit{Hubble} Space Telescope (HST) recently completed an unprecedented number of orbits as part of a Director's Discretionary Time (DDT) project called Ultraviolet Legacy Library of Young Stars as Essential Standards (ULLYSES\footnote{\url{https://ullyses.stsci.edu}}; \citealt{Roman-Duval2020}). The ULLYSES project is targeting over 250 massive stars in the SMC and LMC, with the HST data consisting of high-quality COS and STIS UV spectra. To support these data, an ESO large program, X-Shooting ULLYSES (XShootU\footnote{\url{https://massivestars.org/xshootu/}}), has assembled medium-resolution optical spectra with the XShooter spectrograph \citep{Vernet2011}. Spectroscopic parameters, in particular effective temperature and luminosity, are available from the first papers from the XShootU collaboration \citep{Vink2023a, Hawcroft2024a}, but are also supplemented by spectroscopic studies by \citet{Mokiem2006a} and \citet{Bouret2013a}.


        \subsection{TESS light curve extraction}
        \label{subsection: TESS}
        
        There are two major challenges to overcome to extract reliable TESS light curves of SMC and LMC stars. We note that the TESS mission was not designed for this purpose, and hence specialised software tools are needed and a moderately high failure rate in this work is inevitable. The first challenge is that distant stars beyond the Milky Way are quite faint for the brightness limit of the TESS mission. Second, such stars are usually contaminated since they suffer from high crowding and blending in the relatively large pixels (21~arcsec) of the TESS CCDs. These limitations have the potential to severely limit astrophysical inference based on light curves extracted using standard simple aperture photometry (SAP) techniques, which is common for Galactic massive stars (see e.g. \citealt{Bowman2022a}). 
        
        To overcome these challenges, we used the {\tt tglc} software package\footnote{\url{https://github.com/TeHanHunter/TESS_Gaia_Light_Curve}}, which allows a user to extract light curves of (faint and blended) TESS targets using an effective point spread function (ePSF) fitting methodology \citep{Han_T_2023a}. This ePSF method has been shown to be effective at disentangling different sources in regions with high crowding because it uses a Bayesian framework with priors for source locations informed by Gaia astrometry \citep{Gaia2016a, Gaia2023a}. We refer to \citet{Han_T_2023a} for full technical and numerical details of the well-documented and open-source {\tt tglc} software tool. We performed testing against other PSF-fitting software tools (e.g. {\tt photutils}; \citealt{photutils_2022}), and concluded that the {\tt tglc} software provides, on average, a more robust light curve extraction, especially when compared to standard SAP approaches applied to SMC and LMC massive stars \citep{VanDaele_MASTER}. 
        
        As a sanity check, we visually inspected all the diagnostic plots provided by the {\tt tglc} software package and the extracted ePSF light curves for all stars. Unsuccessful cases were easily flagged based on the standard deviation of the extracted ePSF light curve exceeding 50~mmag, which is primarily because such a light curve is dominated by systematics or extremely high Poisson noise levels because it contains very little flux from the stellar target and is dominated by background flux. The typical amplitudes of SLF variability for Galactic massive stars are well below this (e.g. \citealt{Bowman2019b, Bowman2020b}), meaning that unsuccessful cases are easily flagged automatically. In general, we were guided by previous studies of Galactic massive stars with TESS (e.g. \citealt{Bowman2020b, Bowman2022a}) in terms of the typical amplitude and period ranges of coherent pulsations and SLF variability in massive stars. Since SMC and LMC massive stars have (much) larger noise levels in their light curves compared to Galactic stars, because they are (much) fainter, we were unable to probe the faintest stars in the XShootU sample, which correspond to the lowest mass stars (i.e. late-B stars). Unfortunately, the majority of the XShootU sample of massive stars are quite faint and/or too heavily blended to extract reliable light curves even with the ePSF methodology. 
        
        To be classified as a successful case, each star had to pass all the quality checks discussed above, but we also checked for consistent signatures of SLF variability in its light curve in at least two consecutive TESS sectors (if available since several SMC stars only have a single TESS sector). Consistent in this context means a star has comparable SLF variability morphology (i.e. the same $\nu_{\rm char}$ value within a $2\sigma$ confidence interval) in at least two TESS sectors. Successful cases were subsequently detrended using a low-order polynomial to remove any remaining long-period instrumental artefacts, which is necessary and typical for TESS mission data (see \citealt{Bowman2020b, Bowman2022a}). From the XShootU sample of SMC and LMC stars, we defined a sub-sample of 23 successfully extracted SMC massive stars that are within or proximate to the sub-surface convection zone stability window including asteroseismically calibrated amounts of CBM (cf. Fig.~\ref{figure: HRD}). Additionally, we were able to extract reliable light curves for 21 LMC massive stars, but these are all outside of their corresponding sub-surface convection zone stability window. The names and spectroscopic parameters of the sub-samples of SMC and LMC massive stars are provided in Tables~\ref{table: SMC stars} and \ref{table: LMC stars}, respectively, and the location of these sub-samples in the HR~diagram are shown in Fig.~\ref{figure: HRD}. We note that the SMC sub-sample includes stars both inside and outside of the sub-surface convection zone stability window, allowing us to investigate differences in the observed SLF variability.


\section{SLF variability in SMC and LMC massive stars}
\label{section: results}

Starting from the XShootU sample, we identified a sub-sample of 23 SMC massive stars that satisfy the following criteria: (i) reliable stellar parameters in the literature; (ii) proximate to or within the sub-surface convection zone stability windows that include CBM calibrated by asteroseismology; and (iii) robust ePSF light curves extracted using the {\tt tglc} software package \citep{Han_T_2023a}. The locations of the SMC and LMC sub-samples in the HR~diagram in relation to our calculated evolution tracks and the corresponding sub-surface convection stability windows are shown in Fig.~\ref{figure: HRD}.

\begin{figure*}
\centering
\includegraphics[width=0.45\textwidth]{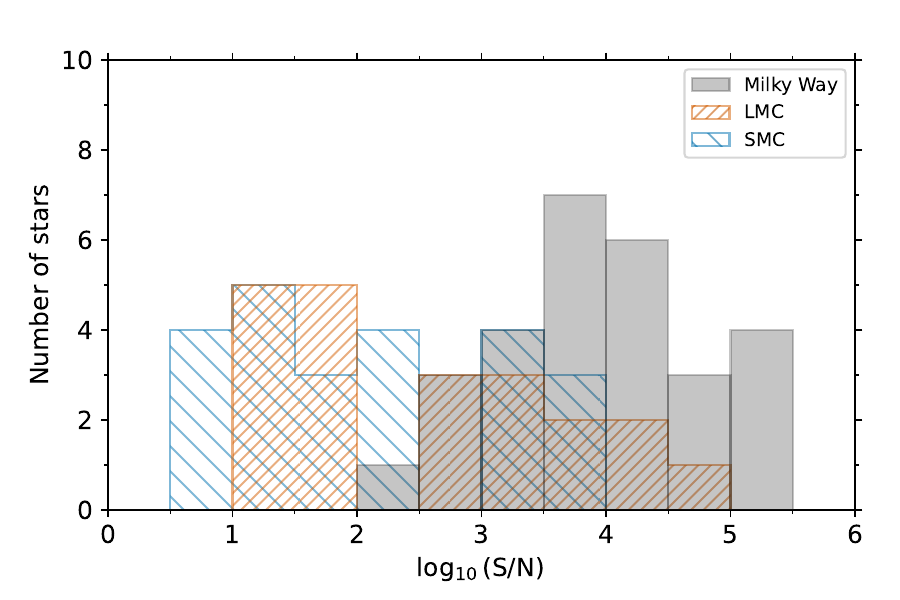}
\includegraphics[width=0.45\textwidth]{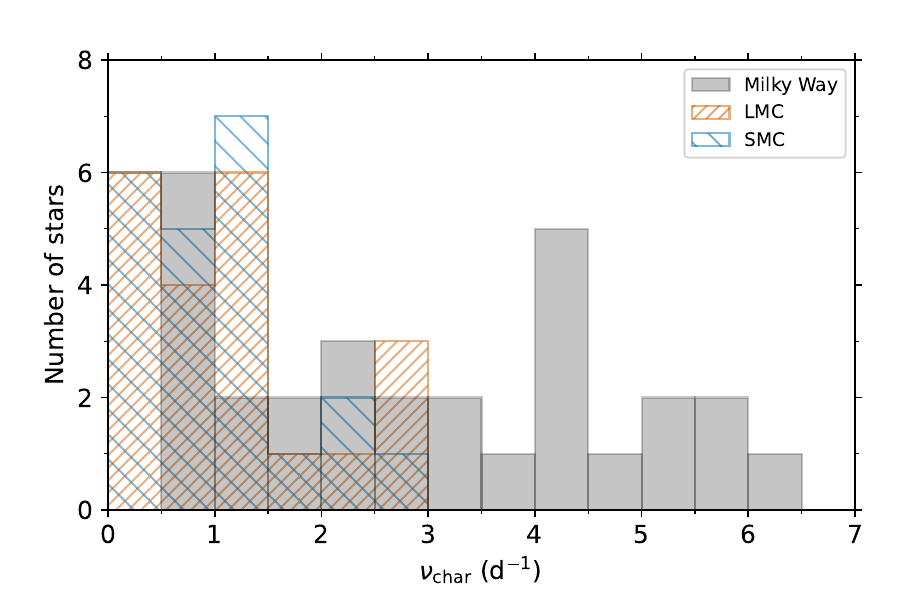}
\includegraphics[width=0.45\textwidth]{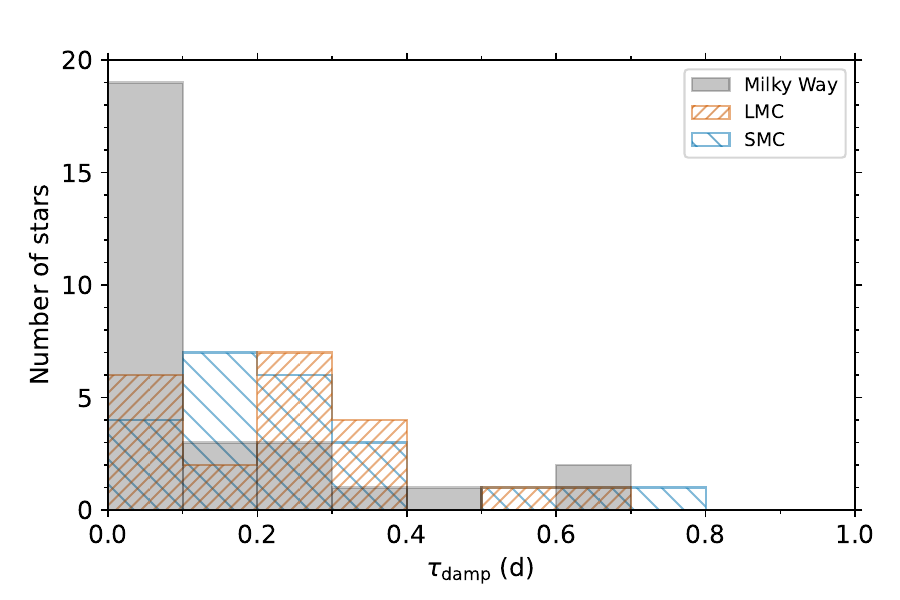}
\includegraphics[width=0.45\textwidth]{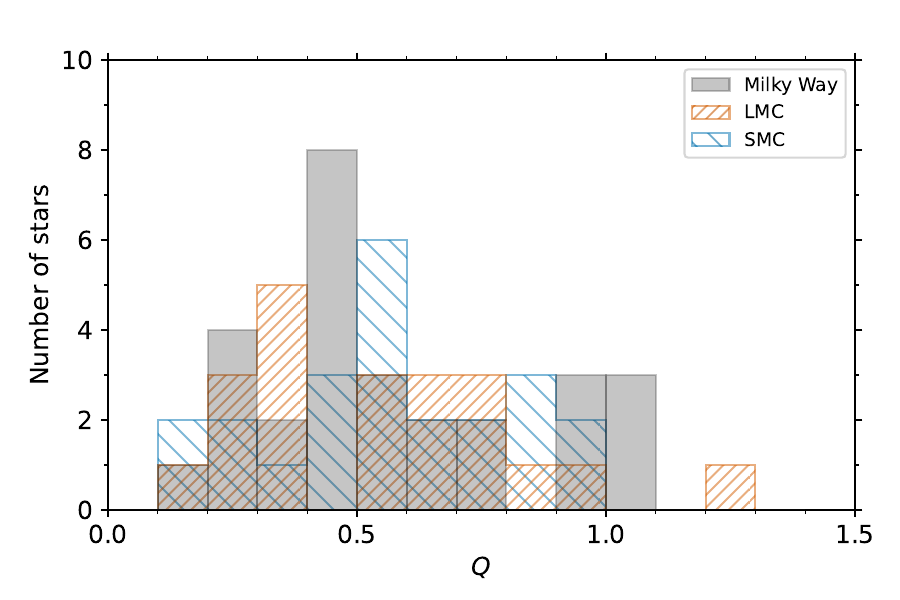}
\caption{Histograms of the signal-to-noise ratio (S/N); characteristic frequency, $\nu_{\rm char}$; damping timescale, $\tau_{\rm damp}$; and the quality factor, $Q$, of SLF variability for the Milky Way,  LMC, and SMC massive stars.}
\label{figure: histograms}
\end{figure*}

We fitted the extracted ePSF light curves using the Gaussian process (GP) regression light curve fitting methodology of \citet{Bowman2022b} based on the {\tt celerite2} software package\footnote{\url{https://celerite2.readthedocs.io/en/latest/}} \citep{Foreman-Mackey2017, Zenodo_celerite_v0.3.1}. Similarly to \citet{Bowman2022b}, who applied this method to Galactic massive stars, we used a damped simple harmonic oscillator (SHO) kernel with a damping timescale, $\tau_{\rm damp}$; amplitude, $\sigma_{A}$; and characteristic variability timescale, $\rho_{\rm damp}$ (from which a characteristic frequency, $\nu_{\rm char}$, can be calculated), as free parameters, all with large uniform priors. An additional term for the jitter, $C_{\rm w}$, was also included, which is related to the amount of white noise in a light curve and is assigned a Gaussian prior with a width corresponding to standard deviation of the light curve (see \citealt{Bowman2022b}). With best-fit parameters determined by the GP regression method, we then estimated the statistical confidence intervals for all GP regression free parameters using a Hamiltonian Monte Carlo with a no U-turn sampler (NUTS) using the {\tt pymc3} software package\footnote{\url{https://docs.pymc.io/en/v3/index.html}} \citep{Salvatier_2016}. The resultant $\nu_{\rm char}$ values and their $2\sigma$ confidence intervals for the sub-samples of SMC and LMC stars are provided in Tables~\ref{table: SMC stars} and \ref{table: LMC stars}, respectively. This numerical analysis methodology was shown to be more efficient and more robust against systematic uncertainties when analysing SLF variability in Galactic massive stars compared to directly fitting an amplitude spectrum with a Lorentzian-like function \citep{Bowman2022b}. 

After fitting the light curves for all stars in our SMC and LMC sub-samples, it is immediately clear that the vast majority of the stars exhibit SLF variability similar to their Galactic counterparts (e.g. \citealt{Bowman2019a, Bowman2019b, Bowman2020b}). This is confirmed using model selection and a Bayesian information criterion (BIC) to invalidate a white noise-only model following a similar approach used by \citet{Bowman2022b}. We recall that the faintest stars in our sample correspond to the lowest mass main-sequence stars in the SMC; we expect these stars to have the smallest amplitudes in their SLF variability, but also the noisiest light curves. Therefore, we demonstrate for the first time that SLF variability is also common across a wide range of metallicity values, as well as mass and age for stars above $\sim$20~M$_{\odot}$. We show the TESS light curves, frequency spectra, and GP regression model for the SMC and LMC sub-samples of stars in Figs.~\ref{figure: SMC 1}--\ref{figure: SMC 3} and \ref{figure: LMC 1}--\ref{figure: LMC 3}, respectively. 

For comparison, we define a third sub-sample of Galactic massive stars based on the best-fitting GP regression models for 30 massive stars in the Milky Way from \citet{Bowman2022b}. We show histograms of the important GP regression fitting parameters for all the massive stars across the Milky Way, LMC and SMC sub-samples in Fig.~\ref{figure: histograms}. Interestingly, the amplitude of the SLF variability is slightly lower on average for the SMC sub-sample of massive stars compared to the LMC. This is achieved by defining a signal-to-noise ratio (S/N) of SLF variability as the ratio of the maximum and minimum values in the power spectral density (PSD) of the GP regression fit. These serve as useful proxies for $\alpha_0$ and $C_{\rm W}$, respectively (cf. \citealt{Bowman2020b}), and allow a comparison of stars with very different white noise contributions in their light curves. The values of PSD$_{\rm max}$ and PSD$_{\rm min}$ are given in Tables~\ref{table: SMC stars} and \ref{table: LMC stars}, for the SMC and LMC sub-samples, respectively. For example, the inferred value of the amplitude of the SLF variability in a frequency spectrum in SMC stars can be artificially higher if inferred from directly fitting a frequency spectrum because of the increased contribution of white noise (i.e. $C_{\rm W}$), which can be large within the sub-sample of SMC stars. With the caveat of it being a small sample of stars to compare, the histograms in Fig.~\ref{figure: histograms} show similar values for S/N; characteristic frequency, $\nu_{\rm char}$; damping timescale, $\tau_{\rm damp}$; and quality factor, $Q$, across the three metallicity regimes.

We find similar morphologies of SLF variability for stars in the SMC regardless of whether they are located inside or outside of their corresponding stability window. However some stars in both the SMC and LMC sub-samples have noisier detections of SLF variability owing to relatively high levels of Poisson (i.e. white) noise in their light curves. All SLF variability fits are statistically significant according to model selection compared to a white noise-only model using the corresponding BIC values (cf. \citealt{Bowman2022b}). Some noisier detections of SLF variability yield (much) larger relative uncertainties in their inferred $\nu_{\rm char}$ values (i.e. of order 100\% or more) and/or low S/N values (i.e. $<100$) in their SLF variability. We note  that such stars remain robust detections of SLF variability, but are not necessarily well-characterised stars in terms of their inferred $\nu_{\rm char}$ values owing to having, on average, noisier light curves.

Notwithstanding the caveat of  lower mass stars having noisier light curves, on average, we note a trend in that SMC and LMC sub-samples of stars closer to the ZAMS having generally smaller S/N values in their SLF variability compared to those closer to the TAMS, albeit with a large amount of scatter. This agrees with the same trend found by \citet{Bowman2020b} in that the amplitude of SLF variability is a function of main-sequence age for Galactic massive stars. This means that the same trend with age is apparent for low-metallicity stars as Galactic massive stars. However, the small sample size, the relatively large errors, and the scatter in this trend mean that we cannot use the amplitude of SLF variability to validate or invalidate different interpretations for the physical mechanism causing SLF variability. On the other hand, this trend does demonstrate that the evolutionary properties of a massive star can be probed from its light curve as shown previously by \citet{Bowman2020b}.

\begin{figure*}
\centering
\includegraphics[width=0.9\textwidth]{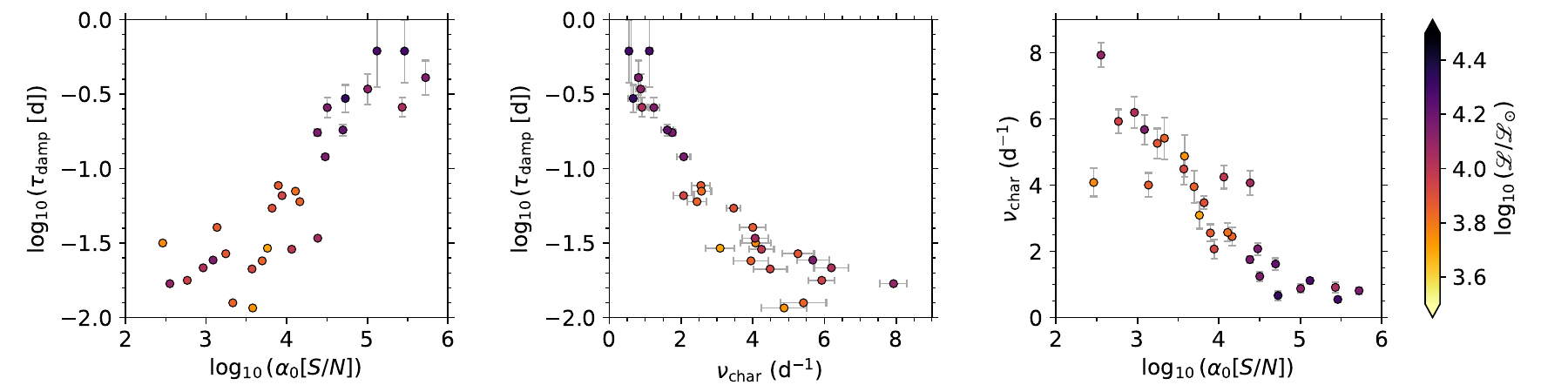}
\includegraphics[width=0.9\textwidth]{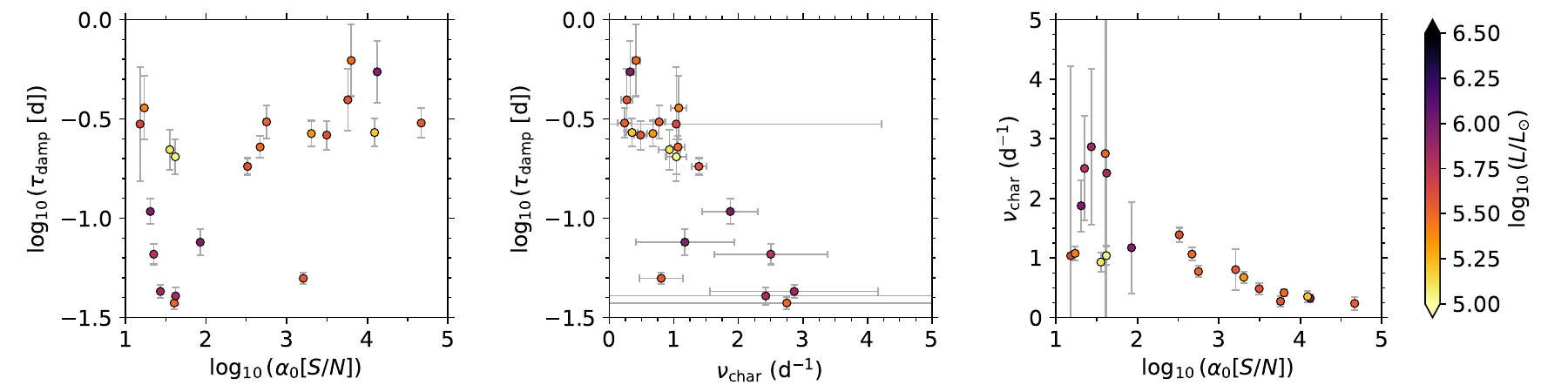}
\includegraphics[width=0.9\textwidth]{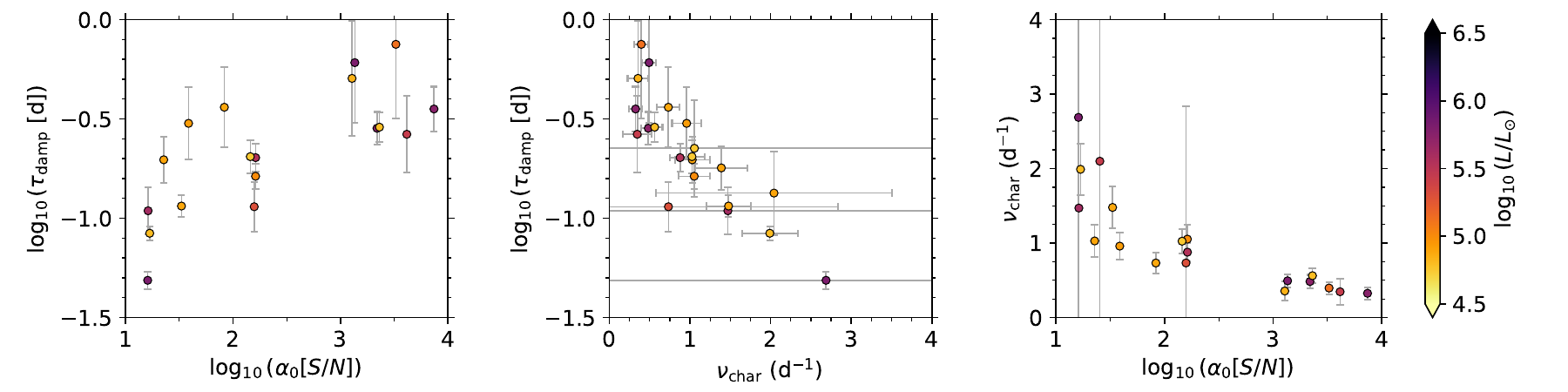}
\caption{Correlations between (spectroscopic) luminosity and SLF variability parameters for the sub-samples of Milky Way, LMC, and SMC massive stars studied in this work are shown (from top to bottom). We note that the spectroscopic effective luminosity is defined as $\mathscr{L} := T_{\rm eff}^{4} / g$ \citep{Langer2014a} with values taken from \citet{Bowman2020b} for consistency. The uncertainties plotted are the $2\sigma$ confidence intervals from the combined methodology of a GP regression and Hamiltonian Monte Carlo with a no U-turn sampler \citep{Bowman2022b}.}
\label{figure: correlations}
\end{figure*}

To further demonstrate the common morphology of SLF variability across all metallicity regimes, the correlations among the S/N; characteristic frequency, $\nu_{\rm char}$; and damping timescale, $\tau_{\rm damp}$, of the SLF variability for the sub-samples of Galactic, LMC, and SMC massive stars studied in this work are colour-coded by luminosity, and are shown in Fig.~\ref{figure: correlations}. For the Milky Way sub-sample stars, we use the spectroscopic effective luminosity with values from \citet{Bowman2020b}, which is defined as $\mathscr{L} := T_{\rm eff}^{4} / g$ \citep{Langer2014a}, due to uncertain values in galactic extinction and reddening and to be consistent with the previous analyses of SLF variability across the HR~diagram in this series of papers. The photometric and GP regression fitting parameters are less precise for the LMC and SMC sub-samples, owing to the noisier TESS light curves, but the same trends in Fig.~\ref{figure: correlations} exist in all three sub-samples. As noted by previous studies, the diversity in SLF variability parameters is high, and could be linked to different rotation rates or inclinations of the stars, which give rise to different surface amplitudes of IGWs (see \citealt{Bowman2022b}). However,  the similar correlations across all metallicity regimes support a common physical origin for SLF variability that is seemingly insensitive to metallicity. For example, massive low-metallicity stars that are luminous have, on average, lower $\nu_{\rm char}$ values and larger SLF variability amplitudes, as previously demonstrated for Galactic massive stars \citep{Bowman2019b, Bowman2020b}. It should be noted that the stars with the noisiest light curves, and thus the largest relative uncertainty in their SLF variability morphology, are typically the faintest stars and are at or beyond the nominal operating brightness limit of TESS around $V \gtrsim 13$~mag. It is challenging to draw strong conclusions on the trends of SLF variability in the HR~diagram based on the currently available sample because the instrumental and astrophysical biases mean fainter stars have lower masses and also have smaller amplitude SLF variability. 

These results lead to two possible conclusions. First, the similar morphology of SLF variability for massive stars in the SMC both inside and outside of the sub-surface convection stability window indicates that SLF variability is unlikely to be exclusively caused by sub-surface convection. This is supported by massive stars having similar morphology in their SLF variability across the metallicity regimes of the Milky Way, LMC, and SMC. However, this assumes that the stability windows (cf. Sect.~\ref{section: method theory}) are based on an accurate implementation of physics in 1D non-rotating structure models, and thus can act as good predictors of observables. A sub-surface stability window does not appear to exist for Milky Way metallicity \citep{Jermyn2022a}, and SLF variability is ubiquitous for such stars \citep{Bowman2019b, Bowman2020b}. Second, if SLF variability is in fact caused by sub-surface convection, these results imply that 1D structure models do not provide reasonable predictors for observables associated with SLF variability. Both of these conclusions may be correct, which emphasises the need to continue to compare observations to hydrodynamical simulations of stellar interiors with 3D spherical geometry and realistic rotation rates (e.g. \citealt{Aerts2015c, Edelmann2019a, Anders_E_2023b}).

On the other hand, since the S/N of SLF variability is somewhat lower for SMC stars compared to LMC and Galactic massive stars, it is a reasonable inference that the observed SLF variability arises from both core convection and sub-surface convection. For LMC and Galactic stars, and their somewhat larger SLF variability amplitudes and their more efficient sub-surface convection zones, the observed SLF variability could be comprised of IGWs from both core convection and sub-surface convection. Instead, for SMC stars with their somewhat smaller SLF variability amplitudes and their less efficient (or absent) sub-surface convection zones, the observed SLF variability could be exclusively caused by IGWs from core convection. Therefore, on balance, these results support the conclusion that 1D structure models cannot be used to rule out core convection as a contributing factor to SLF variability, even if it is arguably not the dominant contributor for Galactic massive stars, which remains to be seen. In addition to metallicity, it is expected from stellar evolution theory and models that sub-surface convection is more important for stars approaching the TAMS given the evolutionary changes to their structures are more conducive to this mechanism than compared to stars near the ZAMS \citep{Bowman2023b}.


\section{Discussion and conclusions}
\label{section: conclusions}

In this work we   demonstrated for the first time that SLF variability is omnipresent in massive stars above 20~M$_{\odot}$ in low-metallicity environments when sufficiently high-quality light curves are available. We achieved this using a sub-sample of massive stars in the SMC and LMC galaxies from the XShootU collaboration \citep{Vink2023a}. We  extracted optimised ePSF light curves from the TESS mission FFI data using the {\tt tglc} software package \citep{Han_T_2023a}, and analysed them using a GP regression fitting methodology \citep{Bowman2022b}. We find that all SMC massive stars have similar SLF variability morphologies to each other, but also to LMC and Galactic massive stars \citep{Bowman2019b, Bowman2020b, Bowman2022b}. Hence, this demonstrates that the physical mechanism(s) responsible for SLF variability in massive stars appears to be largely independent of a star's metallicity, and predominantly set by a star's location in the HR~diagram (i.e. a star's mass and age).

We also find that the morphology of SLF variability in massive stars in the SMC, and in particular its observed amplitude, does not systematically depend on whether a star is within the stability window for sub-surface convection predicted by non-rotating 1D structure models. This is evidenced by SMC massive stars inside and outside of their stability window having similar S/N values in their SLF variability. These stability windows are defined based on the Rayleigh number for convection \citep{Jermyn2022a}, and have been updated in this work to include asteroseismically calibrated constraints on interior mixing \citep{Bowman2020c}. We note that the approach by \citet{Jermyn2022a} to use the Rayleigh number as a definition for the onset of convection produces larger stability windows for SMC stars compared to LMC stars because of their lower metallicities. Owing to the small size of the LMC stability window, it is clear that none of the stars studied in this work fall inside the corresponding sub-surface convection zone stability window even if relatively large confidence intervals for their spectroscopic estimates of effective temperature and luminosity are considered. On the other hand, the fact that the SMC sub-sample includes stars both inside and outside of the corresponding sub-surface convection zone stability window and have similar SLF variability is a challenge to using 1D structure models for predictions of sub-surface convection as a universal explanation for SLF variability. 

Owing to the brightness limit of the TESS mission, our sample with successfully extracted light curves is biased to more massive stars as these are brighter. Moreover, a conclusion on how common SLF variability is between $3 \lesssim M \lesssim 20$~M$_{\odot}$ in the SMC, as previously claimed for Galactic stars by \citet{Bowman2019a} and \citet{Bowman2019b}, cannot be made in this work. The astrophysical bias of our sample is that fainter stars are less massive and if such stars follow the same trend as Galactic stars means they also have smaller amplitude SLF variability, but it is also more difficult to extract light curves for fainter stars. In the future, the ESA PLATO mission \citep{Rauer2024a*} will observe the LMC and SMC with smaller pixels, a longer continuous time span, and an improved photometric precision compared to TESS, allowing us to push to fainter stars.

Important limitations of the 1D structure models used in this work and elsewhere in the literature include the lack of rotation, but also uncertainties in opacity tables for stellar interiors (see discussion by \citealt{Aerts2018b}), which may lead to smaller or larger stability windows dependent on how the physical prescriptions for these processes are incorporated in 1D structure models. For example, \citet{Jermyn2022a} discuss substantially larger stability windows for rotating 1D structure models, even for very slow rotation periods of 1000~d, and a modest further increase for models with rotation periods between 1000~d and 2~d. However, in all cases studied, as expected, rotation has a stabilising effect for convection because it increases the critical Rayleigh number. This means that our non-rotating stability windows are underestimates of the true size of such regions when realistic rotation rates are considered, which only strengthens our conclusions that convection does not play a role in SLF variability for low-metallicity massive stars.

A full parameter study that allows   a range of metallicity values, rotation rates, and chemical mixing processes has not been investigated here and is the subject of future work. For example, rotation breaks the assumption of spherical symmetry and leads to gravity darkening, which would arguably lead to different depths for sub-surface convection as a function of stellar latitude. By extension, this could lead to different SLF variability amplitudes as a function of latitude, meaning one could expect different properties for SLF variability for stars of different inclinations with respect to the observer. Moreover, 1D stellar structure models, including those calculated in this work use mixing-length theory, which has limitations in modelling the true dynamics of convective zones in 3D spherical geometry (see the review by \citealt{Joyce2023b}). There is also growing evidence from modern hydrodynamical simulations that convection is extremely inefficient in the sub-surface convection zones of Galactic massive stars and that radiation is the dominant energy transport mechanism (see e.g. \citealt{Schultz2022a, Schultz2023c, Debnath2024a}). This begs the question of whether these regions should continue to be called sub-surface convection zones.

A complementary result of this work is that we do not find any convincing cases of massive stars with coherent pulsation modes excited by the heat-engine mechanism in the LMC and SMC targets considered. Studies that combine time series spectroscopy (i.e. identifying spectral line profile variability) with photometry have yielded a lower incidence rate of $\beta$~Cep and SPB pulsators and a higher incidence rate of pulsating Be stars in the SMC compared to the galaxy (see e.g. \citealt{Diago2008}). The lower opacities of SMC and LMC massive stars is a natural explanation for a less efficient heat-engine mechanism exciting coherent pulsation modes (e.g. \citealt{Salmon2012}). However, predictions of pulsation excitation mechanism based on 1D stellar structure models typically lack rotation and atomic diffusion including radiative levitation, which have both been shown to be important in exciting coherent pulsation modes in Galactic main-sequence B-type stars \citep{Townsend2005e, Szewczuk2017a, Rehm2024a}. Interestingly, the reduced incidence of coherent pulsators in the literature and lack of coherent pulsators in our sample further supports the conclusion of weaker (or absent) sub-surface convection zones for low-metallicity stars. Therefore, a natural next step on the theoretical side is to include rotation and atomic diffusion including radiative levitation in the calculations of sub-surface convection zone stability windows for low-metallicity massive stars.

The physical mechanism behind SLF variability in massive stars is debated in the literature, but all currently proposed mechanisms have advantages and limitations when confronted with observations (see the review by \citealt{Bowman2023b}). A much larger sample of SMC massive stars with accurate stellar parameters is expected in the near future and will be highly advantageous to explore the physical mechanisms responsible for SLF variability, which is incorporated within the XShootU \citep{Vink2023a} and BLOeM \citep{Shenar2024a} projects. Regardless, our proof-of-concept study demonstrates that it is possible to probe the excitation mechanism of SLF variability at low metallicity and constrain its dependence on a star's location in the HR~diagram. Equally, our results also demonstrate the important limitations of using (non-rotating) 1D structure models for producing observables to be compared to observations of SLF variability.

Finally, our study provides strong motivation for hydrodynamicists to perform low-metallicity simulations of massive stars in 3D spherical geometry, and to include realistic rotation rates, since this may elucidate the latitudinal properties of IGWs excited by sub-surface convection and core convection. We postulate that the high diversity in SLF variability morphologies is caused by a combination of unknown inclination angles combined with a wide range of rotational velocities. Thus, a range of latitudinally dependent interior mixing efficiencies may be at work, which are not well calibrated in massive star evolution theory. Specifically, addressing how convection, and also turbulence, generated by a radiation-dominated iron-bump in the envelopes of massive stars produce SLF variability in 3D spherical geometry including rotation is required in order to accurately compare theoretical amplitudes and frequencies of IGWs to observations. In this study the omnipresence of SLF variability across a   wide range of metallicity values suggests that convection within the iron bump may play a minor role, especially for younger and lower metallicity massive stars.  In the future, a much larger sample of massive stars in the SMC and LMC with accurate spectroscopic parameters delivered by the XShootU and BLOeM projects combined with hydrodynamical simulations will allow a new and exciting dimension to be added to the growing interest in massive star variability studies.


\begin{acknowledgements}

The authors thank the referee for useful feedback that improved the clarity of this work, Adam Jermyn for useful discussions and kindly providing well-documented and open-access MESA inlists (\url{https://github.com/adamjermyn/conv_trends}) to compute evolution tracks and stability windows, as well as Daniel Lecoanet and Matteo Cantiello for useful discussions during the `Probes of Transport in Stars' workshop hosted at KITP in 2021. This research was supported in part by the National Science Foundation (NSF) under Grant Number NSF PHY-1748958. DMB gratefully acknowledges funding from the Research Foundation Flanders (FWO) by means of a senior postdoctoral fellowship (grant agreement No. 1286521N), an FWO long stay travel grant (agreement No. V411621N), as well as UK Research and Innovation (UKRI) in the form of a Frontier Research grant under the UK government's ERC Horizon Europe funding guarantee (SYMPHONY; grant number: EP/Y031059/1), and a Royal Society University Research Fellowship (URF; grant number: URF{\textbackslash}R1{\textbackslash}231631). MM gratefully acknowledges funding from FWO by means of a PhD scholarship (11F7120N), and TVR gratefully acknowledges the KU Leuven Research Council (C16/18/005: PARADISE).

The authors thank the TESS science team for the excellent data, which were obtained from the Mikulski Archive for Space Telescopes (MAST) at the Space Telescope Science Institute (STScI; \url{https://archive.stsci.edu/missions-and-data/tess}), which is operated by the Association of Universities for Research in Astronomy, Inc., under NASA contract NAS5-26555. Support to MAST for these data is provided by the NASA Office of Space Science via grant NAG5-7584 and by other grants and contracts. Funding for the TESS mission is provided by the NASA Explorer Program. This research has made use of the SIMBAD database, operated at CDS, Strasbourg, France; the SAO/NASA Astrophysics Data System; and the VizieR catalog access tool, CDS, Strasbourg, France. 

This research has made use of the following software packages: {\tt tglc} \citep{Han_T_2023a} for light curve extraction (\url{https://github.com/TeHanHunter/TESS_Gaia_Light_Curve}), {\tt celerite2} \citep{Foreman-Mackey2017} for GP regression fitting (\url{https://celerite2.readthedocs.io/en/latest/}) and {\tt pymc3} \citep{Salvatier_2016} for confidence interval estimation (\url{https://github.com/pymc-devs/pymc}), as well as {\tt matplotlib} \citep{Matplotlib_2007}, {\tt seaborn} \citep{Seaborn_2021}, and {\tt numpy} \citep{Numpy_2006, Numpy_2011, Numpy_2020}. For the purpose of open access, the authors have applied a CC BY licence to the author accepted manuscript version: \url{https://arxiv.org/abs/2410.12726}. Data products that support the results in this paper are publicly available via the Zenodo repository: \url{https://zenodo.org/records/14018449}.

\end{acknowledgements}


\bibliographystyle{aa}
\bibliography{SLF_SMC_accepted}

\begin{thebibliography}{98}
\expandafter\ifx\csname natexlab\endcsname\relax\def\natexlab#1{#1}\fi

\bibitem[{{Aerts}(2021)}]{Aerts2021a}
{Aerts}, C. 2021, Reviews of Modern Physics, 93, 015001

\bibitem[{{Aerts} {et~al.}(2010){Aerts}, {Christensen-Dalsgaard}, \&
  {Kurtz}}]{ASTERO_Book}
{Aerts}, C., {Christensen-Dalsgaard}, J., \& {Kurtz}, D.~W. 2010,
  Asteroseismology (Springer)

\bibitem[{{Aerts} {et~al.}(2018){Aerts}, {Molenberghs}, {Michielsen},
  {Pedersen}, {Bj{\"o}rklund}, {Johnston}, {Mombarg}, {Bowman}, {Buysschaert},
  {P{\'a}pics}, {Sekaran}, {Sundqvist}, {Tkachenko}, {Truyaert}, {Van Reeth},
  \& {Vermeyen}}]{Aerts2018b}
{Aerts}, C., {Molenberghs}, G., {Michielsen}, M., {et~al.} 2018, \apjs, 237, 15

\bibitem[{{Aerts} \& {Rogers}(2015)}]{Aerts2015c}
{Aerts}, C. \& {Rogers}, T.~M. 2015, \apjl, 806, L33

\bibitem[{{Aerts} {et~al.}(2003){Aerts}, {Thoul}, {Daszy{\'n}ska}, {Scuflaire},
  {Waelkens}, {Dupret}, {Niemczura}, \& {Noels}}]{Aerts2003d}
{Aerts}, C., {Thoul}, A., {Daszy{\'n}ska}, J., {et~al.} 2003, Science, 300,
  1926

\bibitem[{{Anders} {et~al.}(2023){Anders}, {Lecoanet}, {Cantiello}, {Burns},
  {Hyatt}, {Kaufman}, {Townsend}, {Brown}, {Vasil}, {Oishi}, \&
  {Jermyn}}]{Anders_E_2023b}
{Anders}, E.~H., {Lecoanet}, D., {Cantiello}, M., {et~al.} 2023, Nature
  Astronomy, 7, 1228

\bibitem[{{Auvergne} {et~al.}(2009){Auvergne}, {Bodin}, {Boisnard}, {Buey},
  {Chaintreuil}, {Epstein}, {Jouret}, {Lam-Trong}, {Levacher}, {Magnan},
  {Perez}, {Plasson}, {Plesseria}, {Peter}, {Steller}, {Tiph{\`e}ne}, {Baglin},
  {Agogu{\'e}}, {Appourchaux}, {Barbet}, {Beaufort}, {Bellenger}, {Berlin},
  {Bernardi}, {Blouin}, {Boumier}, {Bonneau}, {Briet}, {Butler}, {Cautain},
  {Chiavassa}, {Costes}, {Cuvilho}, {Cunha-Parro}, {de Oliveira Fialho},
  {Decaudin}, {Defise}, {Djalal}, {Docclo}, {Drummond}, {Dupuis}, {Exil},
  {Faur{\'e}}, {Gaboriaud}, {Gamet}, {Gavalda}, {Grolleau}, {Gueguen},
  {Guivarc'h}, {Guterman}, {Hasiba}, {Huntzinger}, {Hustaix}, {Imbert},
  {Jeanville}, {Johlander}, {Jorda}, {Journoud}, {Karioty}, {Kerjean},
  {Lafond}, {Lapeyrere}, {Landiech}, {Larqu{\'e}}, {Laudet}, {Le Merrer},
  {Leporati}, {Leruyet}, {Levieuge}, {Llebaria}, {Martin}, {Mazy}, {Mesnager},
  {Michel}, {Moalic}, {Monjoin}, {Naudet}, {Neukirchner}, {Nguyen-Kim},
  {Ollivier}, {Orcesi}, {Ottacher}, {Oulali}, {Parisot}, {Perruchot},
  {Piacentino}, {Pinheiro da Silva}, {Platzer}, {Pontet}, {Pradines},
  {Quentin}, {Rohbeck}, {Rolland}, {Rollenhagen}, {Romagnan}, {Russ}, {Samadi},
  {Schmidt}, {Schwartz}, {Sebbag}, {Smit}, {Sunter}, {Tello}, {Toulouse},
  {Ulmer}, {Vandermarcq}, {Vergnault}, {Wallner}, {Waultier}, \&
  {Zanatta}}]{Auvergne2009}
{Auvergne}, M., {Bodin}, P., {Boisnard}, L., {et~al.} 2009, \aap, 506, 411

\bibitem[{{Blomme} {et~al.}(2011){Blomme}, {Mahy}, {Catala}, {Cuypers},
  {Gosset}, {Godart}, {Montalban}, {Ventura}, {Rauw}, {Morel}, {Degroote},
  {Aerts}, {Noels}, {Michel}, {Baudin}, {Baglin}, {Auvergne}, \&
  {Samadi}}]{Blomme2011b}
{Blomme}, R., {Mahy}, L., {Catala}, C., {et~al.} 2011, \aap, 533, A4

\bibitem[{{Borucki} {et~al.}(2010){Borucki}, {Koch}, {Basri}, {Batalha},
  {Brown}, {Caldwell}, {Caldwell}, {Christensen-Dalsgaard}, {Cochran},
  {DeVore}, {Dunham}, {Dupree}, {Gautier}, {Geary}, {Gilliland}, {Gould},
  {Howell}, {Jenkins}, {Kondo}, {Latham}, {Marcy}, {Meibom}, {Kjeldsen},
  {Lissauer}, {Monet}, {Morrison}, {Sasselov}, {Tarter}, {Boss}, {Brownlee},
  {Owen}, {Buzasi}, {Charbonneau}, {Doyle}, {Fortney}, {Ford}, {Holman},
  {Seager}, {Steffen}, {Welsh}, {Rowe}, {Anderson}, {Buchhave}, {Ciardi},
  {Walkowicz}, {Sherry}, {Horch}, {Isaacson}, {Everett}, {Fischer}, {Torres},
  {Johnson}, {Endl}, {MacQueen}, {Bryson}, {Dotson}, {Haas}, {Kolodziejczak},
  {Van Cleve}, {Chandrasekaran}, {Twicken}, {Quintana}, {Clarke}, {Allen},
  {Li}, {Wu}, {Tenenbaum}, {Verner}, {Bruhweiler}, {Barnes}, \&
  {Prsa}}]{Borucki2010}
{Borucki}, W.~J., {Koch}, D., {Basri}, G., {et~al.} 2010, Science, 327, 977

\bibitem[{{Bouret} {et~al.}(2013){Bouret}, {Lanz}, {Martins}, {Marcolino},
  {Hillier}, {Depagne}, \& {Hubeny}}]{Bouret2013a}
{Bouret}, J.~C., {Lanz}, T., {Martins}, F., {et~al.} 2013, \aap, 555, A1

\bibitem[{{Bowman}(2020)}]{Bowman2020c}
{Bowman}, D.~M. 2020, Frontiers in Astronomy and Space Sciences, 7, 70

\bibitem[{{Bowman}(2023)}]{Bowman2023b}
{Bowman}, D.~M. 2023, \apss, 368, 107

\bibitem[{{Bowman} {et~al.}(2019{\natexlab{a}}){Bowman}, {Aerts}, {Johnston},
  {Pedersen}, {Rogers}, {Edelmann}, {Sim{\'o}n-D{\'{\i}}az}, {Van Reeth},
  {Buysschaert}, {Tkachenko}, \& {Triana}}]{Bowman2019a}
{Bowman}, D.~M., {Aerts}, C., {Johnston}, C., {et~al.} 2019{\natexlab{a}},
  \aap, 621, A135

\bibitem[{{Bowman} {et~al.}(2019{\natexlab{b}}){Bowman}, {Burssens},
  {Pedersen}, {Johnston}, {Aerts}, {Buysschaert}, {Michielsen}, {Tkachenko},
  {Rogers}, {Edelmann}, {Ratnasingam}, {Sim{\'o}n-D{\'\i}az}, {Castro},
  {Moravveji}, {Pope}, {White}, \& {De Cat}}]{Bowman2019b}
{Bowman}, D.~M., {Burssens}, S., {Pedersen}, M.~G., {et~al.}
  2019{\natexlab{b}}, Nature Astronomy, 3, 760

\bibitem[{{Bowman} {et~al.}(2020){Bowman}, {Burssens}, {Sim{\'o}n-D{\'\i}az},
  {Edelmann}, {Rogers}, {Horst}, {R{\"o}pke}, \& {Aerts}}]{Bowman2020b}
{Bowman}, D.~M., {Burssens}, S., {Sim{\'o}n-D{\'\i}az}, S., {et~al.} 2020,
  \aap, 640, A36

\bibitem[{{Bowman} \& {Dorn-Wallenstein}(2022)}]{Bowman2022b}
{Bowman}, D.~M. \& {Dorn-Wallenstein}, T.~Z. 2022, \aap, 668, A134

\bibitem[{{Bowman} {et~al.}(2022){Bowman}, {Vandenbussche}, {Sana},
  {Tkachenko}, {Raskin}, {Delabie}, {Vandoren}, {Royer}, {Garcia}, {Van Reeth},
  \& {CubeSpec Collaboration}}]{Bowman2022a}
{Bowman}, D.~M., {Vandenbussche}, B., {Sana}, H., {et~al.} 2022, \aap, 658, A96

\bibitem[{{Bradley} {et~al.}(2022){Bradley}, {Sip{\H o}cz}, {Robitaille},
  {Tollerud}, {Vin{\'\i}cius}, {Deil}, {Barbary}, {Wilson}, {Busko}, {Donath},
  {G{\"u}nther}, {Cara}, {Lim}, {Me{\ss}linger}, {Conseil}, A., {Droettboom},
  {Bray}, {Bratholm}, {Barentsen}, {Craig}, {Rathi}, {Pascual}, {Perren},
  {Georgiev}, {de Val-Borro}, {Kerzendorf}, {Bach}, {Quint}, \&
  {Souchereau}}]{photutils_2022}
{Bradley}, L., {Sip{\H o}cz}, B., {Robitaille}, T., {et~al.} 2022,
  astropy/photutils: 1.5.0

\bibitem[{{Briquet} {et~al.}(2011){Briquet}, {Aerts}, {Baglin}, {Nieva},
  {Degroote}, {Przybilla}, {Noels}, {Schiller}, {Vu{\v c}kovi{\'c}}, {Oreiro},
  {Smolders}, {Auvergne}, {Baudin}, {Catala}, {Michel}, \&
  {Samadi}}]{Briquet2011}
{Briquet}, M., {Aerts}, C., {Baglin}, A., {et~al.} 2011, \aap, 527, A112

\bibitem[{{Briquet} {et~al.}(2012){Briquet}, {Neiner}, {Aerts}, {Morel},
  {Mathis}, {Reese}, {Lehmann}, {Costero}, {Echevarria}, {Handler}, {Kambe},
  {Hirata}, {Masuda}, {Wright}, {Yang}, {Pintado}, {Mkrtichian}, {Lee}, {Han},
  {Bruch}, {De Cat}, {Uytterhoeven}, {Lefever}, {Vanautgaerden}, {de Batz},
  {Fr{\'e}mat}, {Henrichs}, {Geers}, {Martayan}, {Hubert}, {Thizy}, \&
  {Tijani}}]{Briquet2012}
{Briquet}, M., {Neiner}, C., {Aerts}, C., {et~al.} 2012, \mnras, 427, 483

\bibitem[{{Burssens} {et~al.}(2023){Burssens}, {Bowman}, {Michielsen},
  {Sim{\'o}n-D{\'\i}az}, {Aerts}, {Vanlaer}, {Banyard}, {Nardetto}, {Townsend},
  {Handler}, {Mombarg}, {Vanderspek}, \& {Ricker}}]{Burssens2023a}
{Burssens}, S., {Bowman}, D.~M., {Michielsen}, M., {et~al.} 2023, Nature
  Astronomy, 7, 913

\bibitem[{{Burssens} {et~al.}(2020){Burssens}, {Sim{\'o}n-D{\'\i}az}, {Bowman},
  {Holgado}, {Michielsen}, {de Burgos}, {Castro}, {Barb{\'a}}, \&
  {Aerts}}]{Burssens2020a}
{Burssens}, S., {Sim{\'o}n-D{\'\i}az}, S., {Bowman}, D.~M., {et~al.} 2020,
  \aap, 639, A81

\bibitem[{{Cantiello} {et~al.}(2009){Cantiello}, {Langer}, {Brott}, {de Koter},
  {Shore}, {Vink}, {Voegler}, {Lennon}, \& {Yoon}}]{Cantiello2009a}
{Cantiello}, M., {Langer}, N., {Brott}, I., {et~al.} 2009, \aap, 499, 279

\bibitem[{{Cantiello} {et~al.}(2021){Cantiello}, {Lecoanet}, {Jermyn}, \&
  {Grassitelli}}]{Cantiello2021b}
{Cantiello}, M., {Lecoanet}, D., {Jermyn}, A.~S., \& {Grassitelli}, L. 2021,
  \apj, 915, 112

\bibitem[{{Debnath} {et~al.}(2024){Debnath}, {Sundqvist}, {Moens}, {Van der
  Sijpt}, {Verhamme}, \& {Poniatowski}}]{Debnath2024a}
{Debnath}, D., {Sundqvist}, J.~O., {Moens}, N., {et~al.} 2024, \aap, 684, A177

\bibitem[{{Degroote} {et~al.}(2009){Degroote}, {Aerts}, {Ollivier}, {Miglio},
  {Debosscher}, {Cuypers}, {Briquet}, {Montalb{\'a}n}, {Thoul}, {Noels}, {De
  Cat}, {Balaguer-N{\'u}{\~n}ez}, {Maceroni}, {Ribas}, {Auvergne}, {Baglin},
  {Deleuil}, {Weiss}, {Jorda}, {Baudin}, \& {Samadi}}]{Degroote2009b}
{Degroote}, P., {Aerts}, C., {Ollivier}, M., {et~al.} 2009, \aap, 506, 471

\bibitem[{{Diago} {et~al.}(2008){Diago}, {Guti{\'e}rrez-Soto}, {Fabregat}, \&
  {Martayan}}]{Diago2008}
{Diago}, P.~D., {Guti{\'e}rrez-Soto}, J., {Fabregat}, J., \& {Martayan}, C.
  2008, \aap, 480, 179

\bibitem[{{Dupret} {et~al.}(2004){Dupret}, {Thoul}, {Scuflaire},
  {Daszy{\'n}ska-Daszkiewicz}, {Aerts}, {Bourge}, {Waelkens}, \&
  {Noels}}]{Dupret2004b}
{Dupret}, M.-A., {Thoul}, A., {Scuflaire}, R., {et~al.} 2004, \aap, 415, 251

\bibitem[{{Dziembowski} {et~al.}(1993){Dziembowski}, {Moskalik}, \&
  {Pamyatnykh}}]{Dziembowski1993f}
{Dziembowski}, W.~A., {Moskalik}, P., \& {Pamyatnykh}, A.~A. 1993, \mnras, 265,
  588

\bibitem[{{Dziembowski} \& {Pamyatnykh}(1993)}]{Dziembowski1993e}
{Dziembowski}, W.~A. \& {Pamyatnykh}, A.~A. 1993, \mnras, 262, 204

\bibitem[{{Dziembowski} \& {Pamyatnykh}(2008)}]{Dziembowski2008}
{Dziembowski}, W.~A. \& {Pamyatnykh}, A.~A. 2008, \mnras, 385, 2061

\bibitem[{{Edelmann} {et~al.}(2019){Edelmann}, {Ratnasingam}, {Pedersen},
  {Bowman}, {Prat}, \& {Rogers}}]{Edelmann2019a}
{Edelmann}, P.~V.~F., {Ratnasingam}, R.~P., {Pedersen}, M.~G., {et~al.} 2019,
  \apj, 876, 4

\bibitem[{{Foreman-Mackey} {et~al.}(2017){Foreman-Mackey}, {Agol},
  {Ambikasaran}, \& {Angus}}]{Foreman-Mackey2017}
{Foreman-Mackey}, D., {Agol}, E., {Ambikasaran}, S., \& {Angus}, R. 2017, \aj,
  154, 220

\bibitem[{{Foreman-Mackey} {et~al.}(2019){Foreman-Mackey}, {Agol}, {Angus},
  {Brewer}, {Austin}, {Meierjurgen Farr}, {Guillochon}, {Czekala}, \&
  {Casey}}]{Zenodo_celerite_v0.3.1}
{Foreman-Mackey}, D., {Agol}, E., {Angus}, R., {et~al.} 2019, {dfm/celerite:
  celerite v0.3.1}

\bibitem[{{Gaia Collaboration} {et~al.}(2016){Gaia Collaboration}, {Prusti},
  {de Bruijne}, {Brown}, {Vallenari}, {Babusiaux}, {Bailer-Jones}, {Bastian},
  {Biermann}, {Evans}, \& et~al.}]{Gaia2016a}
{Gaia Collaboration}, {Prusti}, T., {de Bruijne}, J.~H.~J., {et~al.} 2016,
  \aap, 595, A1

\bibitem[{{Gaia Collaboration} {et~al.}(2023){Gaia Collaboration}, {Vallenari},
  {Brown}, {Prusti}, {de Bruijne}, {Arenou}, {Babusiaux}, {Biermann},
  {Creevey}, {Ducourant}, {Evans}, {Eyer}, {Guerra}, {Hutton}, {Jordi},
  {Klioner}, {Lammers}, {Lindegren}, {Luri}, {Mignard}, {Panem}, {Pourbaix},
  {Randich}, {Sartoretti}, {Soubiran}, {Tanga}, {Walton}, {Bailer-Jones},
  {Bastian}, {Drimmel}, {Jansen}, {Katz}, {Lattanzi}, {van Leeuwen}, {Bakker},
  {Cacciari}, {Casta{\~n}eda}, {De Angeli}, {Fabricius}, {Fouesneau},
  {Fr{\'e}mat}, {Galluccio}, {Guerrier}, {Heiter}, {Masana}, {Messineo},
  {Mowlavi}, {Nicolas}, {Nienartowicz}, {Pailler}, {Panuzzo}, {Riclet}, {Roux},
  {Seabroke}, {Sordo}, {Th{\'e}venin}, {Gracia-Abril}, {Portell}, {Teyssier},
  {Altmann}, {Andrae}, {Audard}, {Bellas-Velidis}, {Benson}, {Berthier},
  {Blomme}, {Burgess}, {Busonero}, {Busso}, {C{\'a}novas}, {Carry}, {Cellino},
  {Cheek}, {Clementini}, {Damerdji}, {Davidson}, {de Teodoro}, {Nu{\~n}ez
  Campos}, {Delchambre}, {Dell'Oro}, {Esquej}, {Fern{\'a}ndez-Hern{\'a}ndez},
  {Fraile}, {Garabato}, {Garc{\'\i}a-Lario}, {Gosset}, {Haigron}, {Halbwachs},
  {Hambly}, {Harrison}, {Hern{\'a}ndez}, {Hestroffer}, {Hodgkin}, {Holl},
  {Jan{\ss}en}, {Jevardat de Fombelle}, {Jordan}, {Krone-Martins}, {Lanzafame},
  {L{\"o}ffler}, {Marchal}, {Marrese}, {Moitinho}, {Muinonen}, {Osborne},
  {Pancino}, {Pauwels}, {Recio-Blanco}, {Reyl{\'e}}, {Riello}, {Rimoldini},
  {Roegiers}, {Rybizki}, {Sarro}, {Siopis}, {Smith}, {Sozzetti}, {Utrilla},
  {van Leeuwen}, {Abbas}, {{\'A}brah{\'a}m}, {Abreu Aramburu}, {Aerts},
  {Aguado}, {Ajaj}, {Aldea-Montero}, {Altavilla}, {{\'A}lvarez}, {Alves},
  {Anders}, {Anderson}, {Anglada Varela}, {Antoja}, {Baines}, {Baker},
  {Balaguer-N{\'u}{\~n}ez}, {Balbinot}, {Balog}, {Barache}, {Barbato},
  {Barros}, {Barstow}, {Bartolom{\'e}}, {Bassilana}, {Bauchet}, {Becciani},
  {Bellazzini}, {Berihuete}, {Bernet}, {Bertone}, {Bianchi}, {Binnenfeld},
  {Blanco-Cuaresma}, {Blazere}, {Boch}, {Bombrun}, {Bossini}, {Bouquillon},
  {Bragaglia}, {Bramante}, {Breedt}, {Bressan}, {Brouillet}, {Brugaletta},
  {Bucciarelli}, {Burlacu}, {Butkevich}, {Buzzi}, {Caffau}, {Cancelliere},
  {Cantat-Gaudin}, {Carballo}, {Carlucci}, {Carnerero}, {Carrasco},
  {Casamiquela}, {Castellani}, {Castro-Ginard}, {Chaoul}, {Charlot}, {Chemin},
  {Chiaramida}, {Chiavassa}, {Chornay}, {Comoretto}, {Contursi}, {Cooper},
  {Cornez}, {Cowell}, {Crifo}, {Cropper}, {Crosta}, {Crowley}, {Dafonte},
  {Dapergolas}, {David}, {David}, {de Laverny}, {De Luise}, {De March}, {De
  Ridder}, {de Souza}, {de Torres}, {del Peloso}, {del Pozo}, {Delbo},
  {Delgado}, {Delisle}, {Demouchy}, {Dharmawardena}, {Di Matteo}, {Diakite},
  {Diener}, {Distefano}, {Dolding}, {Edvardsson}, {Enke}, {Fabre}, {Fabrizio},
  {Faigler}, {Fedorets}, {Fernique}, {Fienga}, {Figueras}, {Fournier},
  {Fouron}, {Fragkoudi}, {Gai}, {Garcia-Gutierrez}, {Garcia-Reinaldos},
  {Garc{\'\i}a-Torres}, {Garofalo}, {Gavel}, {Gavras}, {Gerlach}, {Geyer},
  {Giacobbe}, {Gilmore}, {Girona}, {Giuffrida}, {Gomel}, {Gomez},
  {Gonz{\'a}lez-N{\'u}{\~n}ez}, {Gonz{\'a}lez-Santamar{\'\i}a},
  {Gonz{\'a}lez-Vidal}, {Granvik}, {Guillout}, {Guiraud},
  {Guti{\'e}rrez-S{\'a}nchez}, {Guy}, {Hatzidimitriou}, {Hauser}, {Haywood},
  {Helmer}, {Helmi}, {Sarmiento}, {Hidalgo}, {Hilger}, {H{\l}adczuk}, {Hobbs},
  {Holland}, {Huckle}, {Jardine}, {Jasniewicz}, {Jean-Antoine Piccolo},
  {Jim{\'e}nez-Arranz}, {Jorissen}, {Juaristi Campillo}, {Julbe}, {Karbevska},
  {Kervella}, {Khanna}, {Kontizas}, {Kordopatis}, {Korn}, {K{\'o}sp{\'a}l},
  {Kostrzewa-Rutkowska}, {Kruszy{\'n}ska}, {Kun}, {Laizeau}, {Lambert},
  {Lanza}, {Lasne}, {Le Campion}, {Lebreton}, {Lebzelter}, {Leccia}, {Leclerc},
  {Lecoeur-Taibi}, {Liao}, {Licata}, {Lindstr{\o}m}, {Lister}, {Livanou},
  {Lobel}, {Lorca}, {Loup}, {Madrero Pardo}, {Magdaleno Romeo}, {Managau},
  {Mann}, {Manteiga}, {Marchant}, {Marconi}, {Marcos}, {Marcos Santos},
  {Mar{\'\i}n Pina}, {Marinoni}, {Marocco}, {Marshall}, {Martin Polo},
  {Mart{\'\i}n-Fleitas}, {Marton}, {Mary}, {Masip}, {Massari},
  {Mastrobuono-Battisti}, {Mazeh}, {McMillan}, {Messina}, {Michalik}, {Millar},
  {Mints}, {Molina}, {Molinaro}, {Moln{\'a}r}, {Monari}, {Mongui{\'o}},
  {Montegriffo}, {Montero}, {Mor}, {Mora}, {Morbidelli}, {Morel}, {Morris},
  {Muraveva}, {Murphy}, {Musella}, {Nagy}, {Noval}, {Oca{\~n}a}, {Ogden},
  {Ordenovic}, {Osinde}, {Pagani}, {Pagano}, {Palaversa}, {Palicio},
  {Pallas-Quintela}, {Panahi}, {Payne-Wardenaar}, {Pe{\~n}alosa Esteller},
  {Penttil{\"a}}, {Pichon}, {Piersimoni}, {Pineau}, {Plachy}, {Plum}, {Poggio},
  {Pr{\v{s}}a}, {Pulone}, {Racero}, {Ragaini}, {Rainer}, {Raiteri}, {Rambaux},
  {Ramos}, {Ramos-Lerate}, {Re Fiorentin}, {Regibo}, {Richards}, {Rios Diaz},
  {Ripepi}, {Riva}, {Rix}, {Rixon}, {Robichon}, {Robin}, {Robin}, {Roelens},
  {Rogues}, {Rohrbasser}, {Romero-G{\'o}mez}, {Rowell}, {Royer}, {Ruz Mieres},
  {Rybicki}, {Sadowski}, {S{\'a}ez N{\'u}{\~n}ez}, {Sagrist{\`a} Sell{\'e}s},
  {Sahlmann}, {Salguero}, {Samaras}, {Sanchez Gimenez}, {Sanna},
  {Santove{\~n}a}, {Sarasso}, {Schultheis}, {Sciacca}, {Segol}, {Segovia},
  {S{\'e}gransan}, {Semeux}, {Shahaf}, {Siddiqui}, {Siebert}, {Siltala},
  {Silvelo}, {Slezak}, {Slezak}, {Smart}, {Snaith}, {Solano}, {Solitro},
  {Souami}, {Souchay}, {Spagna}, {Spina}, {Spoto}, {Steele},
  {Steidelm{\"u}ller}, {Stephenson}, {S{\"u}veges}, {Surdej}, {Szabados},
  {Szegedi-Elek}, {Taris}, {Taylor}, {Teixeira}, {Tolomei}, {Tonello}, {Torra},
  {Torra}, {Torralba Elipe}, {Trabucchi}, {Tsounis}, {Turon}, {Ulla}, {Unger},
  {Vaillant}, {van Dillen}, {van Reeven}, {Vanel}, {Vecchiato}, {Viala},
  {Vicente}, {Voutsinas}, {Weiler}, {Wevers}, {Wyrzykowski}, {Yoldas}, {Yvard},
  {Zhao}, {Zorec}, {Zucker}, \& {Zwitter}}]{Gaia2023a}
{Gaia Collaboration}, {Vallenari}, A., {Brown}, A.~G.~A., {et~al.} 2023, \aap,
  674, A1

\bibitem[{{Han} \& {Brandt}(2023)}]{Han_T_2023a}
{Han}, T. \& {Brandt}, T.~D. 2023, \aj, 165, 71

\bibitem[{Harris {et~al.}(2020)Harris, Millman, van~der Walt, Gommers,
  Virtanen, Cournapeau, Wieser, Taylor, Berg, Smith, Kern, Picus, Hoyer, van
  Kerkwijk, Brett, Haldane, del R{\'{i}}o, Wiebe, Peterson,
  G{\'{e}}rard-Marchant, Sheppard, Reddy, Weckesser, Abbasi, Gohlke, \&
  Oliphant}]{Numpy_2020}
Harris, C.~R., Millman, K.~J., van~der Walt, S.~J., {et~al.} 2020, Nature, 585,
  357

\bibitem[{{Hawcroft} {et~al.}(2024){Hawcroft}, {Sana}, {Mahy}, {Sundqvist}, {de
  Koter}, {Crowther}, {Bestenlehner}, {Brands}, {David-Uraz}, {Decin}, {Erba},
  {Garcia}, {Hamann}, {Herrero}, {Ignace}, {Kee}, {Kub{\'a}tov{\'a}},
  {Lefever}, {Moffat}, {Najarro}, {Oskinova}, {Pauli}, {Prinja}, {Puls},
  {Sander}, {Shenar}, {St-Louis}, {ud-Doula}, \& {Vink}}]{Hawcroft2024a}
{Hawcroft}, C., {Sana}, H., {Mahy}, L., {et~al.} 2024, \aap, 688, A105

\bibitem[{{Herwig} {et~al.}(2023){Herwig}, {Woodward}, {Mao}, {Thompson},
  {Denissenkov}, {Lau}, {Blouin}, {Andrassy}, \& {Paul}}]{Herwig2023a}
{Herwig}, F., {Woodward}, P.~R., {Mao}, H., {et~al.} 2023, \mnras, 525, 1601

\bibitem[{{Horst} {et~al.}(2020){Horst}, {Edelmann}, {Andr{\'a}ssy},
  {R{\"o}pke}, {Bowman}, {Aerts}, \& {Ratnasingam}}]{Horst2020a}
{Horst}, L., {Edelmann}, P.~V.~F., {Andr{\'a}ssy}, R., {et~al.} 2020, \aap,
  641, A18

\bibitem[{{Howell} {et~al.}(2014){Howell}, {Sobeck}, {Haas}, {Still},
  {Barclay}, {Mullally}, {Troeltzsch}, {Aigrain}, {Bryson}, {Caldwell},
  {Chaplin}, {Cochran}, {Huber}, {Marcy}, {Miglio}, {Najita}, {Smith},
  {Twicken}, \& {Fortney}}]{Howell2014}
{Howell}, S.~B., {Sobeck}, C., {Haas}, M., {et~al.} 2014, \pasp, 126, 398

\bibitem[{{Hunter} {et~al.}(2008){Hunter}, {Brott}, {Lennon}, {Langer},
  {Dufton}, {Trundle}, {Smartt}, {de Koter}, {Evans}, \&
  {Ryans}}]{Hunter_I_2008b}
{Hunter}, I., {Brott}, I., {Lennon}, D.~J., {et~al.} 2008, \apjl, 676, L29

\bibitem[{{Hunter}(2007)}]{Matplotlib_2007}
{Hunter}, J.~D. 2007, Computing in Science and Engineering, 9, 90

\bibitem[{{Jenkins} {et~al.}(2016){Jenkins}, {Twicken}, {McCauliff},
  {Campbell}, {Sanderfer}, {Lung}, {Mansouri-Samani}, {Girouard}, {Tenenbaum},
  {Klaus}, {Smith}, {Caldwell}, {Chacon}, {Henze}, {Heiges}, {Latham},
  {Morgan}, {Swade}, {Rinehart}, \& {Vanderspek}}]{Jenkins2016b}
{Jenkins}, J.~M., {Twicken}, J.~D., {McCauliff}, S., {et~al.} 2016, in
  \procspie, Vol. 9913, Software and Cyberinfrastructure for Astronomy IV,
  99133E

\bibitem[{{Jermyn} {et~al.}(2022){Jermyn}, {Anders}, \&
  {Cantiello}}]{Jermyn2022a}
{Jermyn}, A.~S., {Anders}, E.~H., \& {Cantiello}, M. 2022, \apj, 926, 221

\bibitem[{{Jermyn} {et~al.}(2023){Jermyn}, {Bauer}, {Schwab}, {Farmer}, {Ball},
  {Bellinger}, {Dotter}, {Joyce}, {Marchant}, {Mombarg}, {Wolf}, {Sunny Wong},
  {Cinquegrana}, {Farrell}, {Smolec}, {Thoul}, {Cantiello}, {Herwig}, {Toloza},
  {Bildsten}, {Townsend}, \& {Timmes}}]{Jermyn2023a}
{Jermyn}, A.~S., {Bauer}, E.~B., {Schwab}, J., {et~al.} 2023, \apjs, 265, 15

\bibitem[{{Joyce} \& {Tayar}(2023)}]{Joyce2023b}
{Joyce}, M. \& {Tayar}, J. 2023, Galaxies, 11, 75

\bibitem[{{Koch} {et~al.}(2010){Koch}, {Borucki}, {Basri}, {Batalha}, {Brown},
  {Caldwell}, {Christensen-Dalsgaard}, {Cochran}, {DeVore}, {Dunham},
  {Gautier}, {Geary}, {Gilliland}, {Gould}, {Jenkins}, {Kondo}, {Latham},
  {Lissauer}, {Marcy}, {Monet}, {Sasselov}, {Boss}, {Brownlee}, {Caldwell},
  {Dupree}, {Howell}, {Kjeldsen}, {Meibom}, {Morrison}, {Owen}, {Reitsema},
  {Tarter}, {Bryson}, {Dotson}, {Gazis}, {Haas}, {Kolodziejczak}, {Rowe}, {Van
  Cleve}, {Allen}, {Chandrasekaran}, {Clarke}, {Li}, {Quintana}, {Tenenbaum},
  {Twicken}, \& {Wu}}]{Koch2010}
{Koch}, D.~G., {Borucki}, W.~J., {Basri}, G., {et~al.} 2010, \apjl, 713, L79

\bibitem[{{Krti{\v c}ka} \& {Feldmeier}(2018)}]{Krticka2018e}
{Krti{\v c}ka}, J. \& {Feldmeier}, A. 2018, \aap, 617, A121

\bibitem[{{Krti{\v{c}}ka} \& {Feldmeier}(2021)}]{Krticka2021b}
{Krti{\v{c}}ka}, J. \& {Feldmeier}, A. 2021, \aap, 648, A79

\bibitem[{{Langer} \& {Kudritzki}(2014)}]{Langer2014a}
{Langer}, N. \& {Kudritzki}, R.~P. 2014, \aap, 564, A52

\bibitem[{{Le Saux} {et~al.}(2023){Le Saux}, {Baraffe}, {Guillet}, {Vlaykov},
  {Morison}, {Pratt}, {Constantino}, \& {Goffrey}}]{LeSaux2023a}
{Le Saux}, A., {Baraffe}, I., {Guillet}, T., {et~al.} 2023, \mnras, 522, 2835

\bibitem[{{Lecoanet} {et~al.}(2022){Lecoanet}, {Bowman}, \& {Van
  Reeth}}]{Lecoanet2022a}
{Lecoanet}, D., {Bowman}, D.~M., \& {Van Reeth}, T. 2022, \mnras, 512, L16

\bibitem[{{Lecoanet} {et~al.}(2021){Lecoanet}, {Cantiello}, {Anders},
  {Quataert}, {Couston}, {Bouffard}, {Favier}, \& {Le Bars}}]{Lecoanet2021a}
{Lecoanet}, D., {Cantiello}, M., {Anders}, E.~H., {et~al.} 2021, \mnras, 508,
  132

\bibitem[{{Lecoanet} {et~al.}(2019){Lecoanet}, {Cantiello}, {Quataert},
  {Couston}, {Burns}, {Pope}, {Jermyn}, {Favier}, \& {Le Bars}}]{Lecoanet2019a}
{Lecoanet}, D., {Cantiello}, M., {Quataert}, E., {et~al.} 2019, \apjl, 886, L15

\bibitem[{{Michielsen} {et~al.}(2021){Michielsen}, {Aerts}, \&
  {Bowman}}]{Michielsen2021a}
{Michielsen}, M., {Aerts}, C., \& {Bowman}, D.~M. 2021, \aap, 650, A175

\bibitem[{{Miglio} {et~al.}(2007){Miglio}, {Montalb{\'a}n}, \&
  {Dupret}}]{Miglio2007a}
{Miglio}, A., {Montalb{\'a}n}, J., \& {Dupret}, M.-A. 2007, \mnras, 375, L21

\bibitem[{{Mokiem} {et~al.}(2007){Mokiem}, {de Koter}, {Evans}, {Puls},
  {Smartt}, {Crowther}, {Herrero}, {Langer}, {Lennon}, {Najarro}, {Villamariz},
  \& {Vink}}]{Mokiem2007a}
{Mokiem}, M.~R., {de Koter}, A., {Evans}, C.~J., {et~al.} 2007, \aap, 465, 1003

\bibitem[{{Mokiem} {et~al.}(2006){Mokiem}, {de Koter}, {Evans}, {Puls},
  {Smartt}, {Crowther}, {Herrero}, {Langer}, {Lennon}, {Najarro}, {Villamariz},
  \& {Yoon}}]{Mokiem2006a}
{Mokiem}, M.~R., {de Koter}, A., {Evans}, C.~J., {et~al.} 2006, \aap, 456, 1131

\bibitem[{Oliphant(2006)}]{Numpy_2006}
Oliphant, T.~E. 2006, A guide to NumPy, Vol.~1 (Trelgol Publishing USA)

\bibitem[{{Pamyatnykh} {et~al.}(2004){Pamyatnykh}, {Handler}, \&
  {Dziembowski}}]{Pamyat2004}
{Pamyatnykh}, A.~A., {Handler}, G., \& {Dziembowski}, W.~A. 2004, \mnras, 350,
  1022

\bibitem[{{Paxton} {et~al.}(2011){Paxton}, {Bildsten}, {Dotter}, {Herwig},
  {Lesaffre}, \& {Timmes}}]{Paxton2011}
{Paxton}, B., {Bildsten}, L., {Dotter}, A., {et~al.} 2011, \apjs, 192, 3

\bibitem[{{Paxton} {et~al.}(2013){Paxton}, {Cantiello}, {Arras}, {Bildsten},
  {Brown}, {Dotter}, {Mankovich}, {Montgomery}, {Stello}, {Timmes}, \&
  {Townsend}}]{Paxton2013}
{Paxton}, B., {Cantiello}, M., {Arras}, P., {et~al.} 2013, \apjs, 208, 4

\bibitem[{{Paxton} {et~al.}(2015){Paxton}, {Marchant}, {Schwab}, {Bauer},
  {Bildsten}, {Cantiello}, {Dessart}, {Farmer}, {Hu}, {Langer}, {Townsend},
  {Townsley}, \& {Timmes}}]{Paxton2015}
{Paxton}, B., {Marchant}, P., {Schwab}, J., {et~al.} 2015, \apjs, 220, 15

\bibitem[{{Paxton} {et~al.}(2018){Paxton}, {Schwab}, {Bauer}, {Bildsten},
  {Blinnikov}, {Duffell}, {Farmer}, {Goldberg}, {Marchant}, {Sorokina},
  {Thoul}, {Townsend}, \& {Timmes}}]{Paxton2018}
{Paxton}, B., {Schwab}, J., {Bauer}, E.~B., {et~al.} 2018, \apjs, 234, 34

\bibitem[{{Paxton} {et~al.}(2019){Paxton}, {Smolec}, {Schwab}, {Gautschy},
  {Bildsten}, {Cantiello}, {Dotter}, {Farmer}, {Goldberg}, {Jermyn}, {Kanbur},
  {Marchant}, {Thoul}, {Townsend}, {Wolf}, {Zhang}, \& {Timmes}}]{Paxton2019}
{Paxton}, B., {Smolec}, R., {Schwab}, J., {et~al.} 2019, \apjs, 243, 10

\bibitem[{{Pedersen} {et~al.}(2021){Pedersen}, {Aerts}, {P{\'a}pics},
  {Michielsen}, {Gebruers}, {Rogers}, {Molenberghs}, {Burssens}, {Garcia}, \&
  {Bowman}}]{Pedersen2021a}
{Pedersen}, M.~G., {Aerts}, C., {P{\'a}pics}, P.~I., {et~al.} 2021, Nature
  Astronomy, 5, 715

\bibitem[{{Ram{\'\i}rez-Agudelo} {et~al.}(2013){Ram{\'\i}rez-Agudelo},
  {Sim{\'o}n-D{\'\i}az}, {Sana}, {de Koter}, {Sab{\'\i}n-Sanjul{\'\i}an}, {de
  Mink}, {Dufton}, {Gr{\"a}fener}, {Evans}, {Herrero}, {Langer}, {Lennon},
  {Ma{\'\i}z Apell{\'a}niz}, {Markova}, {Najarro}, {Puls}, {Taylor}, \&
  {Vink}}]{Ramirez-Agudelo2013}
{Ram{\'\i}rez-Agudelo}, O.~H., {Sim{\'o}n-D{\'\i}az}, S., {Sana}, H., {et~al.}
  2013, \aap, 560, A29

\bibitem[{{Ratnasingam} {et~al.}(2019){Ratnasingam}, {Edelmann}, \&
  {Rogers}}]{Ratnasingam2019a}
{Ratnasingam}, R.~P., {Edelmann}, P.~V.~F., \& {Rogers}, T.~M. 2019, \mnras,
  482, 5500

\bibitem[{{Ratnasingam} {et~al.}(2020){Ratnasingam}, {Edelmann}, \&
  {Rogers}}]{Ratnasingam2020a}
{Ratnasingam}, R.~P., {Edelmann}, P.~V.~F., \& {Rogers}, T.~M. 2020, \mnras,
  497, 4231

\bibitem[{{Ratnasingam} {et~al.}(2023){Ratnasingam}, {Rogers}, {Chowdhury},
  {Handler}, {Vanon}, {Varghese}, \& {Edelmann}}]{Ratnasingam2023a}
{Ratnasingam}, R.~P., {Rogers}, T.~M., {Chowdhury}, S., {et~al.} 2023, \aap,
  674, A134

\bibitem[{{Rauer} {et~al.}(2024){Rauer}, {Aerts}, {Cabrera}, {Deleuil},
  {Erikson}, {Gizon}, {Goupil}, {Heras}, {Lorenzo-Alvarez}, {Marliani},
  {Martin-Garcia}, {Mas-Hesse}, {O'Rourke}, {Osborn}, {Pagano}, {Piotto},
  {Pollacco}, {Ragazzoni}, {Ramsay}, {Udry}, {Appourchaux}, {Benz},
  {Brandeker}, {G{\"u}del}, {Janot-Pacheco}, {Kabath}, {Kjeldsen}, {Min},
  {Santos}, {Smith}, {Suarez}, {Werner}, {Aboudan}, {Abreu}, {Acu{\~n}a},
  {Adams}, {Adibekyan}, {Affer}, {Agneray}, {Agnor}, {Aguirre B{\o}rsen-Koch},
  {Ahmed}, {Aigrain}, {Al-Bahlawan}, {Alcacera Gil}, {Alei}, {Alencar},
  {Alexander}, {Alfonso-Garz{\'o}n}, {Alibert}, {Allende Prieto}, {Almeida},
  {Alonso Sobrino}, {Altavilla}, {Althaus}, {Alonso Alvarez Trujillo},
  {Amarsi}, {Ammler-von Eiff}, {Am{\^o}res}, {Andrade}, {Antoniadis-Karnavas},
  {Ant{\'o}nio}, {Aparicio del Moral}, {Appolloni}, {Arena}, {Armstrong},
  {Aroca Aliaga}, {Asplund}, {Audenaert}, {Auricchio}, {Avelino}, {Baeke},
  {Bailli{\'e}}, {Balado}, {Balestra}, {Ball}, {Ballans}, {Ballot}, {Barban},
  {Barbary}, {Barbieri}, {Barcel{\'o} Forteza}, {Barker}, {Barklem}, {Barnes},
  {Barrado Navascues}, {Barragan}, {Baruteau}, {Basu}, {Baudin}, {Baumeister},
  {Bayliss}, {Bazot}, {Beck}, {Bedding}, {Belkacem}, {Bellinger}, {Benatti},
  {Benomar}, {B{\'e}rard}, {Bergemann}, {Bergomi}, {Bernardo}, {Biazzo},
  {Bignamini}, {Bigot}, {Billot}, {Binet}, {Biondi}, {Biondi}, {Birch},
  {Bitsch}, {Bluhm Ceballos}, {B{\'o}di}, {Bogn{\'a}r}, {Boisse}, {Bolmont},
  {Bonanno}, {Bonavita}, {Bonfanti}, {Bonfils}, {Bonito}, {Bonomo},
  {B{\"o}rner}, {Boro Saikia}, {Borreguero Mart{\'\i}n}, {Borsa}, {Borsato},
  {Bossini}, {Bouchy}, {Bou{\'e}}, {Boufleur}, {Boumier}, {Bourrier}, {Bowman},
  {Bozzo}, {Bradley}, {Bray}, {Bressan}, {Breton}, {Brienza}, {Brito}, {Brogi},
  {Brown}, {Brown}, {Brun}, {Bruno}, {Bruns}, {Buchhave}, {Bugnet}, {Buldgen},
  {Burgess}, {Busatta}, {Busso}, {Buzasi}, {Caballero}, {Cabral}, {Calderone},
  {Cameron}, {Cameron}, {Campante}, {Canto Martins}, {Cara}, {Carone},
  {Carrasco}, {Casagrande}, {Casewell}, {Cassisi}, {Castellani}, {Castro},
  {Catala}, {Catal{\'a}n Fern{\'a}ndez}, {Catelan}, {Cegla}, {Cerruti},
  {Cessa}, {Chadid}, {Chaplin}, {Charpinet}, {Chiappini}, {Chiarucci},
  {Chiavassa}, {Chinellato}, {Chirulli}, {Christensen-Dalsgaard}, {Church},
  {Claret}, {Clarke}, {Claudi}, {Clermont}, {Coelho}, {Coelho}, {Cogato},
  {Colom{\'e}}, {Condamin}, {Conseil}, {Corbard}, {Correia}, {Corsaro},
  {Cosentino}, {Costes}, {Cottinelli}, {Covone}, {Creevey}, {Crida},
  {Csizmadia}, {Cunha}, {Curry}, {da Costa}, {da Silva}, {Dalal}, {Damasso},
  {Damiani}, {Damiani}, {Liduina das Chagas}, {Davies}, {Davies}, {Davies},
  {Davison}, {de Almeida}, {de Angeli}, {Cabral de Barros}, {de Castro
  Le{\~a}o}, {Brito de Freitas}, {de Freitas}, {De Martino}, {Renan de
  Medeiros}, {de Paula}, {de Plaa}, {De Ridder}, {Deal}, {Decin}, {Deeg},
  {Degl'Innocenti}, {Deheuvels}, {del Burgo}, {Del Sordo}, {Delgado-Mena},
  {Demangeon}, {Denk}, {Derekas}, {Desidera}, {Dexet}, {Di Criscienzo}, {Di
  Giorgio}, {Di Mauro}, {Diaz Rial}, {D{\'\i}az-Garc{\'\i}a}, {Dima},
  {Dinuzzi}, {Dionatos}, {Distefano}, {do Nascimento}, {Domingo}, {D'Orazi},
  {Dorn}, {Doyle}, {Duarte}, {Ducellier}, {Dumaye}, {Dumusque}, {Dupret},
  {Eggenberger}, {Ehrenreich}, {Eigm{\"u}ller}, {Eising}, {Emilio}, {Eriksson},
  {Ermocida}, {Isidoro Escate Giribaldi}, {Eschen}, {Estrela}, {Evans},
  {Fabbian}, {Fabrizio}, {Faria}, {Farina}, {Farinato}, {Feliz}, {Feltzing},
  {Fenouillet}, {Ferrari}, {Ferraz-Mello}, {Fialho}, {Fienga}, {Figueira},
  {Fiori}, {Flaccomio}, {Focardi}, {Foley}, {Fontignie}, {Ford}, {Fornazier},
  {Forveille}, {Fossati}, {de Marca Franca}, {da Silva}, {Frasca}, {Fridlund},
  {Furlan}, {Gabler}, {Gaido}, {Gallagher}, {Galli}, {Garcia}, {Garc{\'\i}a
  Hern{\'a}ndez}, {Garcia Munoz}, {Garc{\'\i}a-V{\'a}zquez}, {Garrido Haba},
  {Gaulme}, {Gauthier}, {Gehan}, {Gent}, {Georgieva}, {Ghigo}, {Giana}, {Gill},
  {Girardi}, {Giuliatti Winter}, {Giusi}, {Gomes da Silva}, {G{\'o}mez Zazo},
  {Gomez-Lopez}, {Isai Gonz{\'a}lez Hern{\'a}ndez}, {Gonzalez Murillo},
  {Gorius}, {Gouel}, {Goulty}, {Granata}, {Grenfell}, {Grie{\ss}bach},
  {Grolleau}, {Grouffal}, {Grziwa}, {Guarcello}, {Gueguen}, {Guenther},
  {Guilhem}, {Guillerot}, {Guiot}, {Guterman}, {Guti{\'e}rrez},
  {Guti{\'e}rrez-Canales}, {Hagelberg}, {Haldemann}, {Hall}, {Handberg},
  {Harrison}, {Harrison}, {Hasiba}, {Haswell}, {Hatalova}, {Hatzes}, {Haywood},
  {H{\'e}brard}, {Heckes}, {Heiter}, {Hekker}, {Heller}, {Helling},
  {Helminiak}, {Hemsley}, {Heng}, {Hermans}, {Hermes}, {Hidalgo Torres},
  {Hinkel}, {Hobbs}, {Hodgkin}, {Hofmann}, {Hojjatpanah}, {Houdek}, {Huber},
  {Huesler}, {Hui-Bon-Hoa}, {Huygen}, {Huynh}, {Iro}, {Irwin}, {Irwin},
  {Izidoro}, {Jacquinod}, {Emborg Jannsen}, {Janson}, {Jeszenszky}, {Jiang},
  {Jos{\'e} Jimenez Mancebo}, {Jofre}, {Johansen}, {Johnston}, {Jones},
  {Kallinger}, {K{\'a}lm{\'a}n}, {Kanitz}, {Karjalainen}, {Karjalainen},
  {Karoff}, {Kawaler}, {Kawata}, {Keereman}, {Keiderling}, {Kennedy},
  {Kenworthy}, {Kerschbaum}, {Kidger}, {Kiefer}, {Kintziger}, {Kislyakova},
  {Kiss}, {Klagyivik}, {Klahr}, {Klevas}, {Kochukhov}, {K{\"o}hler}, {Kolb},
  {Koncz}, {Korth}, {Kostogryz}, {Kov{\'a}cs}, {Kov{\'a}cs}, {Kozhura},
  {Krivova}, {Ku{\v{c}}inskas}, {Kuhlemann}, {Kupka}, {Laauwen}, {Labiano},
  {Lagarde}, {Laget}, {Laky}, {Lam}, {Lambrechts}, {Lammer}, {Lanza},
  {Lanzafame}, {Lares Martiz}, {Laskar}, {Latter}, {Lavanant}, {Lawrenson},
  {Lazzoni}, {Lebre}, {Lebreton}, {Lecavelier des Etangs}, {Leinhardt},
  {Leleu}, {Lendl}, {Leto}, {Levillain}, {Libert}, {Lichtenberg}, {Ligi},
  {Lignieres}, {Lillo-Box}, {Linsky}, {Scige Liu}, {Loidolt}, {Longval},
  {Lopes}, {Lorenzani}, {Ludwig}, {Lund}, {Sloth Lundkvist}, {Luri},
  {Maceroni}, {Madden}, {Madhusudhan}, {Maggio}, {Magliano}, {Magrin}, {Mahy},
  {Maibaum}, {Malac-Allain}, {Malapert}, {Malavolta}, {Maldonado}, {Mamonova},
  {Manchon}, {Mann}, {Mantovan}, {Marafatto}, {Marconi}, {Mardling}, {Marigo},
  {Marinoni}, {Marques}, {Marques}, {Marrese}, {Marshall}, {Mart{\'\i}nez
  Perales}, {Mary}, {Marzari}, {Masana}, {Mascher}, {Mathis}, {Mathur},
  {Mattiuci Figueiredo}, {Maxted}, {Mazeh}, {Mazevet}, {Mazzei}, {McCormac},
  {McMillan}, {Menou}, {Merle}, {Meru}, {Mesa}, {Messina}, {M{\'e}sz{\'a}ros},
  {Meunier}, {Meunier}, {Micela}, {Michaelis}, {Michel}, {Michielsen},
  {Michtchenko}, {Miglio}, {Miguel}, {Milligan}, {Mirouh}, {Mitchell},
  {Moedas}, {Molendini}, {Moln{\'a}r}, {Mombarg}, {Montalban}, {Montalto},
  {Monteiro}, {Morales}, {Morales-Calderon}, {Morbidelli}, {Mordasini},
  {Moreau}, {Morel}, {Morello}, {Morin}, {Mortier}, {Mosser}, {Mourard},
  {Mousis}, {Moutou}, {Mowlavi}, {Moya}, {Muehlmann}, {Muirhead}, {Munari},
  {Musella}, {Mustill}, {Nardetto}, {Nardiello}, {Narita}, {Nascimbeni},
  {Nash}, {Neiner}, {Nelson}, {Nettelmann}, {Nicolini}, {Nielsen}, {Niemi},
  {Noack}, {Noels-Grotsch}, {Noll}, {Norazman}, {Norton}, {Nsamba}, {Ofir},
  {Ogilvie}, {Olander}, {Olivetto}, {Olofsson}, {Ong}, {Ortolani}, {Oshagh},
  {Ottacher}, {Ottensamer}, {Ouazzani}, {Paardekooper}, {Pace}, {Pajas},
  {Palacios}, {Palandri}, {Palle}, {Paproth}, {Parro}, {Parviainen}, {Granado},
  {Passegger}, {Pastor-Morales}, {P{\"a}tzold}, {Gade Pedersen}, {Pena
  Hidalgo}, {Pepe}, {Pereira}, {Persson}, {Pertenais}, {Peter}, {Petit},
  {Petit}, {Pezzuto}, {Pichierri}, {Pietrinferni}, {Pinheiro}, {Pinsonneault},
  {Plachy}, {Plasson}, {Plez}, {Poppenhaeger}, {Poretti}, {Portaluri},
  {Portell}, {Frederico Porto de Mello}, {Poyatos}, {Pozuelos}, {Prada Moroni},
  {Pricopi}, {Prisinzano}, {Quade}, {Quirrenbach160}, {Rabanal Reina6},
  {Rabello Soares}, {Raimondo}, {Rainer}, {Ram{\'o}n Rod{\'o}n},
  {Ram{\'o}n-Ballesta}, {Ramos Zapata}, {R{\"a}tz}, {Rauterberg}, {Redman},
  {Redmer}, {Reese}, {Regibo}, {Reiners}, {Reinhold}, {Renie}, {Ribas},
  {Ribeiro}, {Pereira Ricciardi}, {Rice}, {Richard}, {Riello}, {Rieutord},
  {Ripepi}, {Rixon}, {Rockstein}, {Rodr{\'\i}guez}, {Rodr{\'\i}guez D{\'\i}az},
  {Rodriguez Garcia}, {Rodriguez-Gomez}, {Roehlly}, {Roig}, {Rojas-Ayala},
  {Rolf}, {Lysgaard R{\o}rsted}, {Rosado}, {Rosotti}, {Roth}, {Roth},
  {Rousseau}, {Roxburgh}, {Roy}, {Royer}, {Ruane}, {Rufini Mastropasqua}, {Ruiz
  de Galarreta}, {Russi}, {Saar}, {Saillenfest}, {Salaris}, {Salmon}, {Saltas},
  {Samadi}, {Samadi}, {Samra}, {Sanches da Silva}, {Andr{\'e}s S{\'a}nchez
  Carrasco}, {Santerne}, {Santoli}, {Santos}, {Sanz Mesa}, {Sarro},
  {Scandariato}, {Sch{\"a}fer}, {Schlafly}, {Schmider}, {Schneider}, {Schou},
  {Schunker}, {J{\"o}rg Schwarzkopf}, {Serenelli}, {Seynaeve}, {Shan},
  {Shapiro}, {Shipman}, {Sicilia}, {Sierra Sanmartin}, {Sigot}, {Silliman},
  {Silvotti}, {Simon}, {Simoyama Napoli}, {Skarka}, {Smalley}, {Smiljanic},
  {Smit}, {Smith}, {Smith}, {Snellen}, {S{\'o}dor}, {Sohl}, {Solanki},
  {Sortino}, {Sousa}, {Southworth}, {Souto}, {Sozzetti}, {Stamatellos},
  {Stassun}, {Steller}, {Stello}, {Stelzer}, {Stiebeler}, {Stokholm},
  {Storelvmo}, {Strassmeier}, {Str{\o}m}, {Strugarek}, {Sulis}, {{\v{S}}vanda},
  {Szabados}, {Szab{\'o}}, {Szab{\'o}}, {Szuszkiewicz}, {Talens}, {Teti},
  {Theisen}, {Th{\'e}venin}, {Thoul}, {Tiphene}, {Titz-Weider}, {Tkachenko},
  {Tomecki}, {Tonfat}, {Tosi}, {Trampedach}, {Traven}, {Triaud}, {Tr{\o}nnes},
  {Tsantaki}, {Tschentscher}, {Turin}, {Tvaruzka}, {Ulmer}, {Ulmer-Moll},
  {Ulusoy}, {Umbriaco}, {Valencia}, {Valentini}, {Valio}, {Valverde Guijarro},
  {Van Eylen}, {Van Grootel}, {van Kempen}, {Van Reeth}, {Van Zelst},
  {Vandenbussche}, {Vasiliou}, {Vasilyev}, {Vaz de Mascarenhas}, {Vazan}, {Vela
  Nunez}, {Nunes Velloso}, {Ventura}, {Ventura}, {Venturini}, {Trallero},
  {Veras}, {Verdugo}, {Verma}, {Vibert}, {Vicanek Martinez}, {Vida}, {Vigan},
  {Villacorta}, {Villaver}, {Villaverde Aparicio}, {Viotto}, {Vorobyov},
  {Vorontsov}, {Wagner}, {Walloschek}, {Walton}, {Walton}, {Wang}, {Waters},
  {Watson}, {Wedemeyer}, {Weeks}, {Weingril}, {Weiss}, {Wendler}, {West},
  {Westerdorff}, {Westphal}, {Wheatley}, {White}, {Whittaker}, {Wickhusen},
  {Wilson}, {Windsor}, {Winter}, {Lykke Winther}, {Winton}, {Witteck},
  {Witzke}, {Woitke}, {Wolter}, {Wuchterl}, {Wyatt}, {Yang}, {Yu}, {Zanmar
  Sanchez}, {Rosa Zapatero Osorio}, {Zechmeister}, {Zhou}, {Ziemke}, \&
  {Zwintz}}]{Rauer2024a*}
{Rauer}, H., {Aerts}, C., {Cabrera}, J., {et~al.} 2024, arXiv e-prints,
  arXiv:2406.05447

\bibitem[{{Rehm} {et~al.}(2024){Rehm}, {Mombarg}, {Aerts}, {Michielsen},
  {Burssens}, \& {Townsend}}]{Rehm2024a}
{Rehm}, R., {Mombarg}, J. S.~G., {Aerts}, C., {et~al.} 2024, \aap, 687, A175

\bibitem[{{Ricker} {et~al.}(2015){Ricker}, {Winn}, {Vanderspek}, {Latham},
  {Bakos}, {Bean}, {Berta-Thompson}, {Brown}, {Buchhave}, {Butler}, {Butler},
  {Chaplin}, {Charbonneau}, {Christensen-Dalsgaard}, {Clampin}, {Deming},
  {Doty}, {De Lee}, {Dressing}, {Dunham}, {Endl}, {Fressin}, {Ge}, {Henning},
  {Holman}, {Howard}, {Ida}, {Jenkins}, {Jernigan}, {Johnson}, {Kaltenegger},
  {Kawai}, {Kjeldsen}, {Laughlin}, {Levine}, {Lin}, {Lissauer}, {MacQueen},
  {Marcy}, {McCullough}, {Morton}, {Narita}, {Paegert}, {Palle}, {Pepe},
  {Pepper}, {Quirrenbach}, {Rinehart}, {Sasselov}, {Sato}, {Seager},
  {Sozzetti}, {Stassun}, {Sullivan}, {Szentgyorgyi}, {Torres}, {Udry}, \&
  {Villasenor}}]{Ricker2015}
{Ricker}, G.~R., {Winn}, J.~N., {Vanderspek}, R., {et~al.} 2015, Journal of
  Astronomical Telescopes, Instruments, and Systems, 1, 014003

\bibitem[{{Rogers}(2015)}]{Rogers2015}
{Rogers}, T.~M. 2015, \apjl, 815, L30

\bibitem[{{Rogers} {et~al.}(2013){Rogers}, {Lin}, {McElwaine}, \&
  {Lau}}]{Rogers2013b}
{Rogers}, T.~M., {Lin}, D.~N.~C., {McElwaine}, J.~N., \& {Lau}, H.~H.~B. 2013,
  \apj, 772, 21

\bibitem[{{Roman-Duval} {et~al.}(2020){Roman-Duval}, {Proffitt}, {Taylor},
  {Monroe}, {Fischer}, {Fischer}, {Fullerton}, {Aloisi}, {Britt}, {Busko},
  {Carlberg}, {De Rosa}, {Jedrzejewski}, {Lockwood}, {Frazer}, {Hernandez},
  {James}, {Oliveira}, {Plesha}, {Riedel}, {Riley}, {Sahnow}, {Sankrit},
  {Shaw}, {Smith}, {Sohn}, {Som}, {Ubeda}, \& {Welty}}]{Roman-Duval2020}
{Roman-Duval}, J., {Proffitt}, C.~R., {Taylor}, J.~M., {et~al.} 2020, Research
  Notes of the American Astronomical Society, 4, 205

\bibitem[{{Salmon} {et~al.}(2012){Salmon}, {Montalb{\'a}n}, {Morel}, {Miglio},
  {Dupret}, \& {Noels}}]{Salmon2012}
{Salmon}, S., {Montalb{\'a}n}, J., {Morel}, T., {et~al.} 2012, \mnras, 422,
  3460

\bibitem[{{Salmon} {et~al.}(2022){Salmon}, {Moyano}, {Eggenberger},
  {Haemmerl{\'e}}, \& {Buldgen}}]{Salmon2022b}
{Salmon}, S.~J.~A.~J., {Moyano}, F.~D., {Eggenberger}, P., {Haemmerl{\'e}}, L.,
  \& {Buldgen}, G. 2022, \aap, 664, L1

\bibitem[{{Salvatier} {et~al.}(2016){Salvatier}, {Wiecki{\^a}}, \&
  {Fonnesbeck}}]{Salvatier_2016}
{Salvatier}, J., {Wiecki{\^a}}, T.~V., \& {Fonnesbeck}, C. 2016, {PyMC3: Python
  probabilistic programming framework}

\bibitem[{{Sana} {et~al.}(2012){Sana}, {de Mink}, {de Koter}, {Langer},
  {Evans}, {Gieles}, {Gosset}, {Izzard}, {Le Bouquin}, \&
  {Schneider}}]{Sana2012b}
{Sana}, H., {de Mink}, S.~E., {de Koter}, A., {et~al.} 2012, Science, 337, 444

\bibitem[{{Schultz} {et~al.}(2022){Schultz}, {Bildsten}, \&
  {Jiang}}]{Schultz2022a}
{Schultz}, W.~C., {Bildsten}, L., \& {Jiang}, Y.-F. 2022, \apjl, 924, L11

\bibitem[{{Schultz} {et~al.}(2023){Schultz}, {Bildsten}, \&
  {Jiang}}]{Schultz2023c}
{Schultz}, W.~C., {Bildsten}, L., \& {Jiang}, Y.-F. 2023, \apjl, 951, L42

\bibitem[{{Shenar} {et~al.}(2024){Shenar}, {Bodensteiner}, {Sana}, {Crowther},
  {Lennon}, {Abdul-Masih}, {Almeida}, {Backs}, {Berlanas}, {Bernini-Peron},
  {Bestenlehner}, {Bowman}, {Bronner}, {Britavskiy}, {de Koter}, {de Mink},
  {Deshmukh}, {Evans}, {Fabry}, {Gieles}, {Gilkis}, {Gonz{\'a}lez-Tor{\`a}},
  {Gr{\"a}fener}, {G{\"o}tberg}, {Hawcroft}, {H{\'e}nault-Brunet}, {Herrero},
  {Holgado}, {Janssens}, {Johnston}, {Josiek}, {Justham}, {Kalari}, {Katabi},
  {Keszthelyi}, {Klencki}, {Kub{\'a}t}, {Kub{\'a}tov{\'a}}, {Langer},
  {Lefever}, {Ludwig}, {Mackey}, {Mahy}, {Ma{\'\i}z Apell{\'a}niz}, {Mandel},
  {Maravelias}, {Marchant}, {Menon}, {Najarro}, {Oskinova}, {O'Grady},
  {Ovadia}, {Patrick}, {Pauli}, {Pawlak}, {Ramachandran}, {Renzo}, {Rocha},
  {Sander}, {Sayada}, {Schneider}, {Schootemeijer}, {Sch{\"o}sser},
  {Sch{\"u}rmann}, {Sen}, {Shahaf}, {Sim{\'o}n-D{\'\i}az}, {Stoop}, {Toonen},
  {Tramper}, {van Loon}, {Valli}, {van Son}, {Vigna-G{\'o}mez},
  {Villase{\~n}or}, {Vink}, {Wang}, \& {Willcox}}]{Shenar2024a}
{Shenar}, T., {Bodensteiner}, J., {Sana}, H., {et~al.} 2024, \aap, 690, A289

\bibitem[{{Szewczuk} \& {Daszy{\'n}ska-Daszkiewicz}(2017)}]{Szewczuk2017a}
{Szewczuk}, W. \& {Daszy{\'n}ska-Daszkiewicz}, J. 2017, \mnras, 469, 13

\bibitem[{{Thompson} {et~al.}(2024){Thompson}, {Herwig}, {Woodward}, {Mao},
  {Denissenkov}, {Bowman}, \& {Blouin}}]{Thompson_W_2024a}
{Thompson}, W., {Herwig}, F., {Woodward}, P.~R., {et~al.} 2024, \mnras, 531,
  1316

\bibitem[{{Townsend}(2005)}]{Townsend2005e}
{Townsend}, R.~H.~D. 2005, \mnras, 364, 573

\bibitem[{{Van Daele}(2023)}]{VanDaele_MASTER}
{Van Daele}, P. 2023, Master's thesis, KU Leuven, Belgium

\bibitem[{{van der Walt} {et~al.}(2011){van der Walt}, {Colbert}, \&
  {Varoquaux}}]{Numpy_2011}
{van der Walt}, S., {Colbert}, S.~C., \& {Varoquaux}, G. 2011, Computing in
  Science Engineering, 13, 22

\bibitem[{{Vanon} {et~al.}(2023){Vanon}, {Edelmann}, {Ratnasingam}, {Varghese},
  \& {Rogers}}]{Vanon2023a}
{Vanon}, R., {Edelmann}, P.~V.~F., {Ratnasingam}, R.~P., {Varghese}, A., \&
  {Rogers}, T.~M. 2023, \apj, 954, 171

\bibitem[{{Varghese} {et~al.}(2023){Varghese}, {Ratnasingam}, {Vanon},
  {Edelmann}, \& {Rogers}}]{Varghese2023a}
{Varghese}, A., {Ratnasingam}, R.~P., {Vanon}, R., {Edelmann}, P.~V.~F., \&
  {Rogers}, T.~M. 2023, \apj, 942, 53

\bibitem[{{Vernet} {et~al.}(2011){Vernet}, {Dekker}, {D'Odorico}, {Kaper},
  {Kjaergaard}, {Hammer}, {Randich}, {Zerbi}, {Groot}, {Hjorth}, {Guinouard},
  {Navarro}, {Adolfse}, {Albers}, {Amans}, {Andersen}, {Andersen}, {Binetruy},
  {Bristow}, {Castillo}, {Chemla}, {Christensen}, {Conconi}, {Conzelmann},
  {Dam}, {de Caprio}, {de Ugarte Postigo}, {Delabre}, {di Marcantonio},
  {Downing}, {Elswijk}, {Finger}, {Fischer}, {Flores}, {Fran{\c{c}}ois},
  {Goldoni}, {Guglielmi}, {Haigron}, {Hanenburg}, {Hendriks}, {Horrobin},
  {Horville}, {Jessen}, {Kerber}, {Kern}, {Kiekebusch}, {Kleszcz}, {Klougart},
  {Kragt}, {Larsen}, {Lizon}, {Lucuix}, {Mainieri}, {Manuputy}, {Martayan},
  {Mason}, {Mazzoleni}, {Michaelsen}, {Modigliani}, {Moehler}, {M{\o}ller},
  {Norup S{\o}rensen}, {N{\o}rregaard}, {P{\'e}roux}, {Patat}, {Pena}, {Pragt},
  {Reinero}, {Rigal}, {Riva}, {Roelfsema}, {Royer}, {Sacco}, {Santin},
  {Schoenmaker}, {Spano}, {Sweers}, {Ter Horst}, {Tintori}, {Tromp}, {van
  Dael}, {van der Vliet}, {Venema}, {Vidali}, {Vinther}, {Vola}, {Winters},
  {Wistisen}, {Wulterkens}, \& {Zacchei}}]{Vernet2011}
{Vernet}, J., {Dekker}, H., {D'Odorico}, S., {et~al.} 2011, \aap, 536, A105

\bibitem[{{Vink} {et~al.}(2023){Vink}, {Mehner}, {Crowther}, {Fullerton},
  {Garcia}, {Martins}, {Morrell}, {Oskinova}, {St-Louis}, {ud-Doula}, {Sander},
  {Sana}, {Bouret}, {Kub{\'a}tov{\'a}}, {Marchant}, {Martins}, {Wofford}, {van
  Loon}, {Grace Telford}, {G{\"o}tberg}, {Bowman}, {Erba}, {Kalari},
  {Abdul-Masih}, {Alkousa}, {Backs}, {Barbosa}, {Berlanas}, {Bernini-Peron},
  {Bestenlehner}, {Blomme}, {Bodensteiner}, {Brands}, {Evans}, {David-Uraz},
  {Driessen}, {Dsilva}, {Geen}, {G{\'o}mez-Gonz{\'a}lez}, {Grassitelli},
  {Hamann}, {Hawcroft}, {Herrero}, {Higgins}, {John Hillier}, {Ignace},
  {Istrate}, {Kaper}, {Kee}, {Kehrig}, {Keszthelyi}, {Klencki}, {de Koter},
  {Kuiper}, {Laplace}, {Larkin}, {Lefever}, {Leitherer}, {Lennon}, {Mahy},
  {Ma{\'\i}z Apell{\'a}niz}, {Maravelias}, {Marcolino}, {McLeod}, {de Mink},
  {Najarro}, {Oey}, {Parsons}, {Pauli}, {Pedersen}, {Prinja}, {Ramachandran},
  {Ram{\'\i}rez-Tannus}, {Sabhahit}, {Schootemeijer}, {Reyero Serantes},
  {Shenar}, {Stringfellow}, {Sudnik}, {Tramper}, \& {Wang}}]{Vink2023a}
{Vink}, J.~S., {Mehner}, A., {Crowther}, P.~A., {et~al.} 2023, \aap, 675, A154

\bibitem[{Waskom(2021)}]{Seaborn_2021}
Waskom, M.~L. 2021, Journal of Open Source Software, 6, 3021

\bibitem[{{Weiss} {et~al.}(2014){Weiss}, {Rucinski}, {Moffat},
  {Schwarzenberg-Czerny}, {Koudelka}, {Grant}, {Zee}, {Kuschnig}, {Mochnacki},
  {Matthews}, {Orleanski}, {Pamyatnykh}, {Pigulski}, {Alves}, {Guedel},
  {Handler}, {Wade}, \& {Zwintz}}]{Weiss2014}
{Weiss}, W.~W., {Rucinski}, S.~M., {Moffat}, A.~F.~J., {et~al.} 2014, \pasp,
  126, 573

\bibitem[{{Weiss} {et~al.}(2021){Weiss}, {Zwintz}, {Kuschnig}, {Handler},
  {Moffat}, {Baade}, {Bowman}, {Granzer}, {Kallinger}, {Koudelka}, {Lovekin},
  {Neiner}, {Pablo}, {Pigulski}, {Popowicz}, {Ramiaramanantsoa}, {Rucinski},
  {Strassmeier}, \& {Wade}}]{Weiss2021a}
{Weiss}, W.~W., {Zwintz}, K., {Kuschnig}, R., {et~al.} 2021, Universe, 7, 199

\bibitem[{{Zwintz} {et~al.}(2024){Zwintz}, {Pigulski}, {Kuschnig}, {Wade},
  {Doherty}, {Earl}, {Lovekin}, {M{\"u}llner}, {Pich{\'e}-Perrier}, {Steindl},
  {Beck}, {Bicz}, {Bowman}, {Handler}, {Pablo}, {Popowicz},
  {R{\'o}{\.z}a{\'n}ski}, {Miko{\l}ajczyk}, {Baade}, {Koudelka}, {Moffat},
  {Neiner}, {Orlea{\'n}ski}, {Smolec}, {Louis}, {Weiss}, {Wenger}, \&
  {Zoc{\l}o{\'n}ska}}]{Zwintz2024a}
{Zwintz}, K., {Pigulski}, A., {Kuschnig}, R., {et~al.} 2024, \aap, 683, A49

\end{thebibliography}


\onecolumn


\begin{appendix}

\section{Additional figures}
\label{appendix: figures}

\begin{figure*}
\centering
\includegraphics[width=0.41\textwidth]{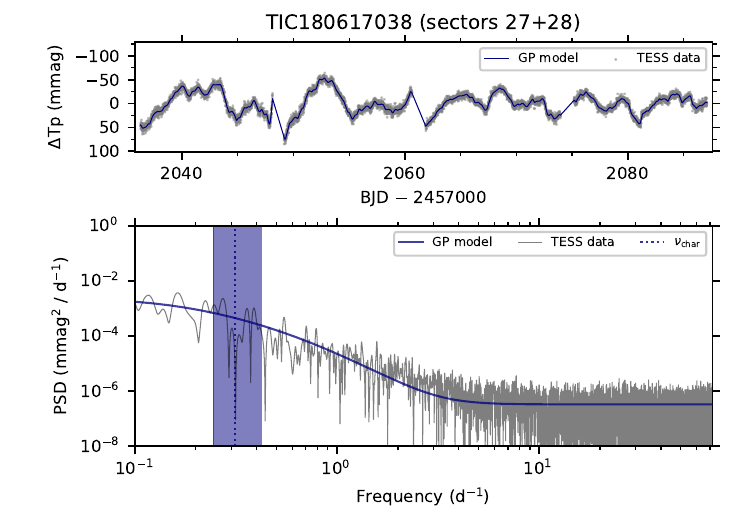}
\includegraphics[width=0.41\textwidth]{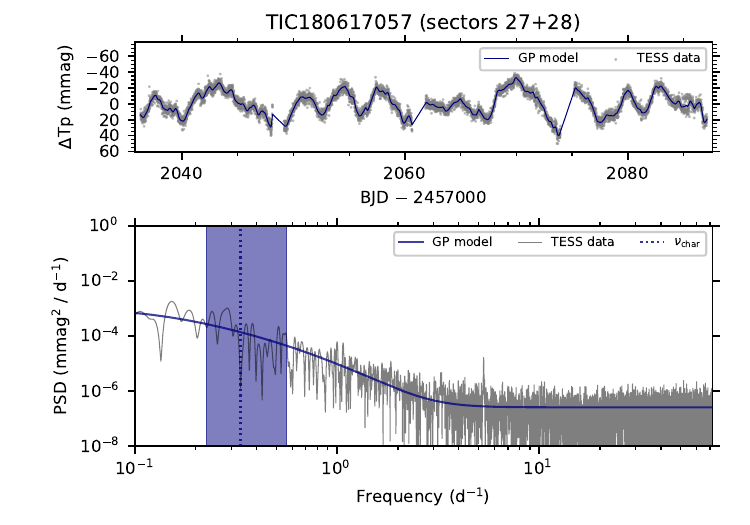}
\includegraphics[width=0.41\textwidth]{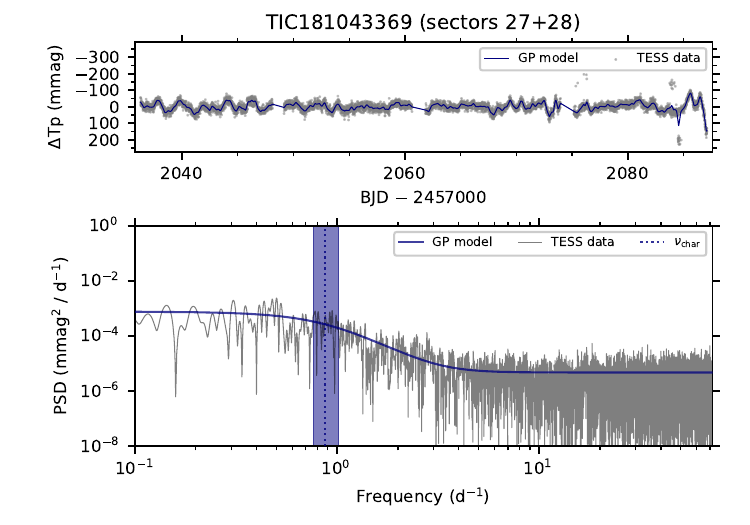}
\includegraphics[width=0.41\textwidth]{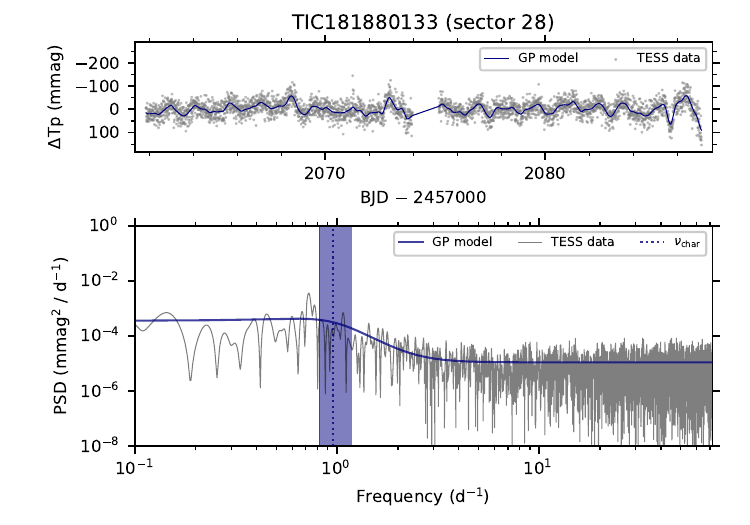}
\includegraphics[width=0.41\textwidth]{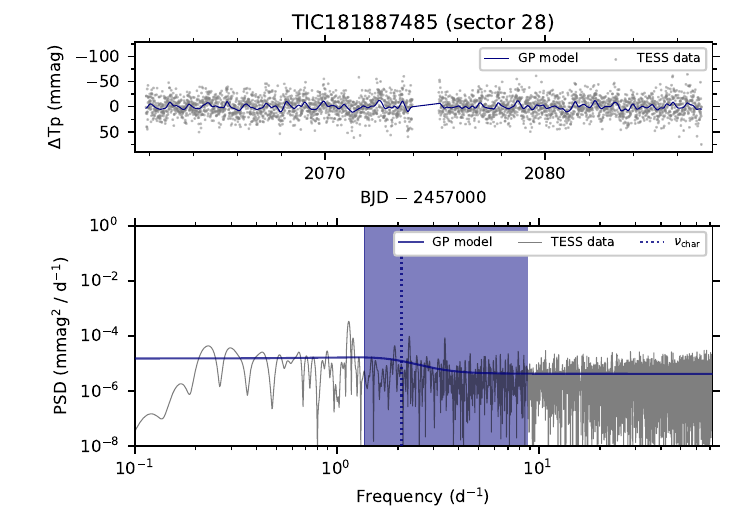}
\includegraphics[width=0.41\textwidth]{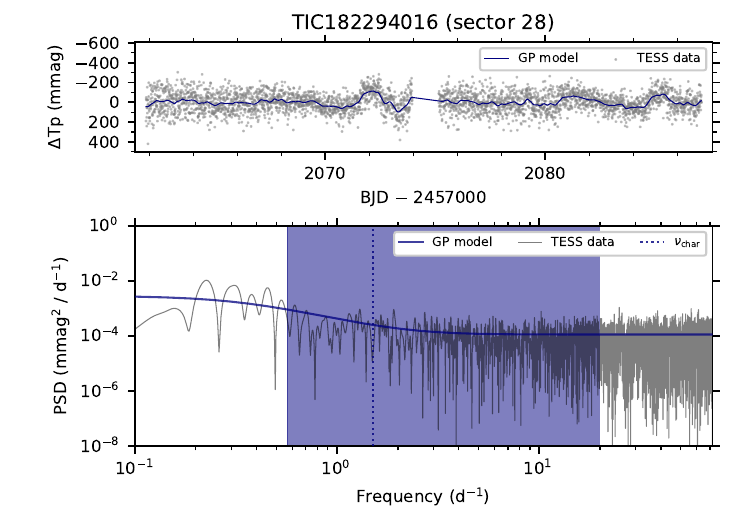}
\includegraphics[width=0.41\textwidth]{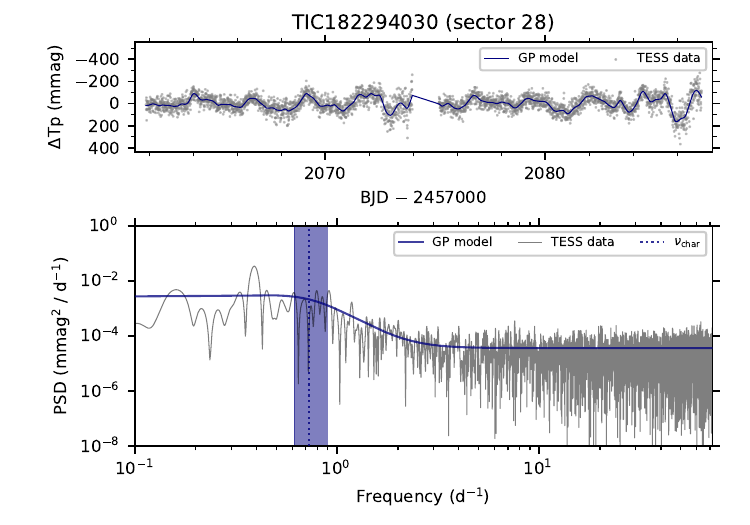}
\includegraphics[width=0.41\textwidth]{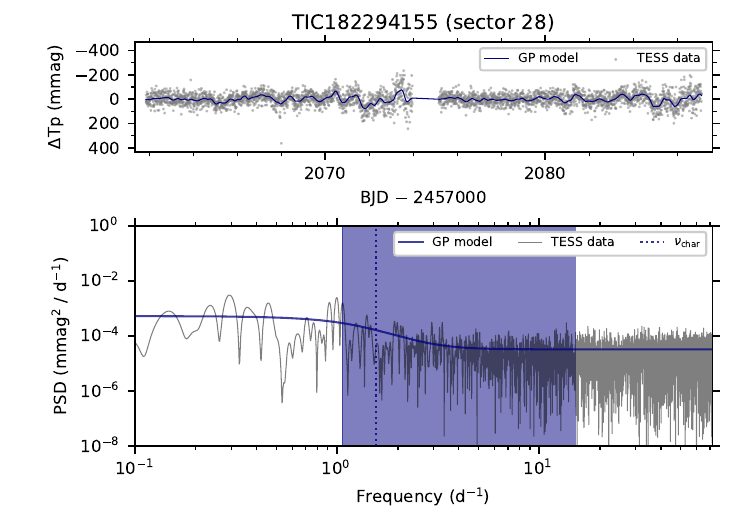}
\caption{Extracted ePSF light curves using {\tt tglc} \citep{Han_T_2023a} and fitted with the GP regression methodology of \citet{Bowman2022b} for SMC massive stars.}
\label{figure: SMC 1}
\end{figure*}

\begin{figure*}
\centering
\includegraphics[width=0.41\textwidth]{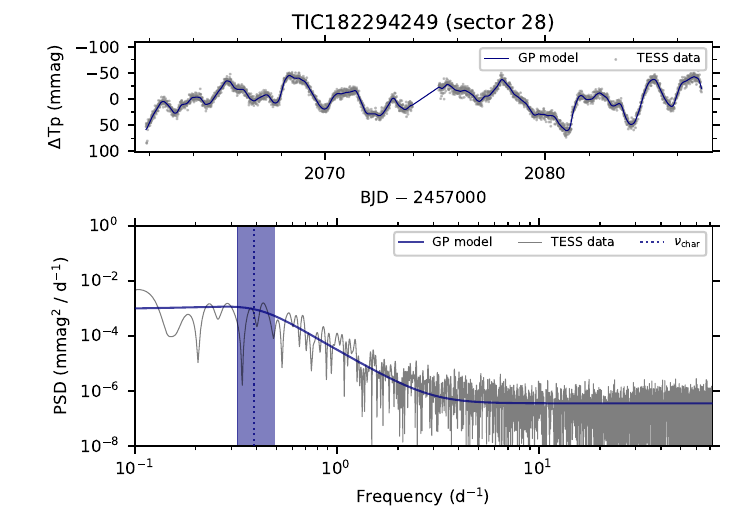}
\includegraphics[width=0.41\textwidth]{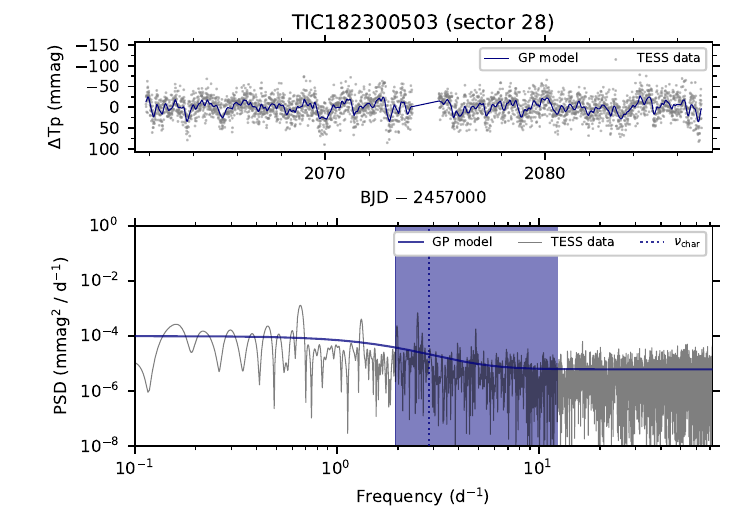}
\includegraphics[width=0.41\textwidth]{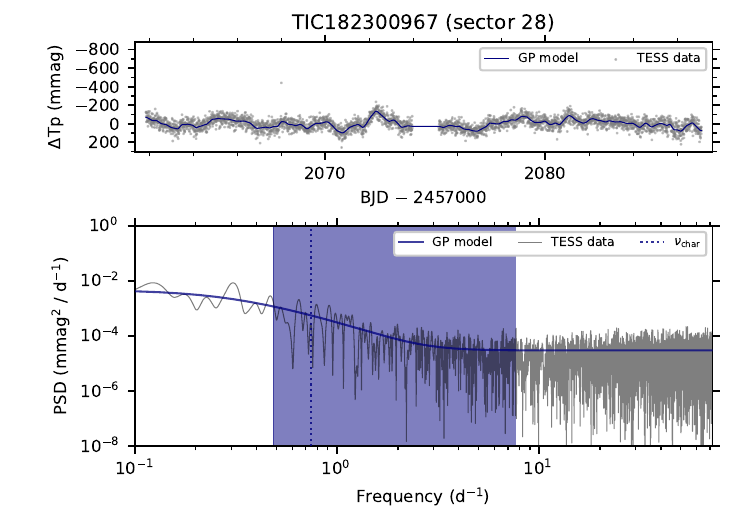}
\includegraphics[width=0.41\textwidth]{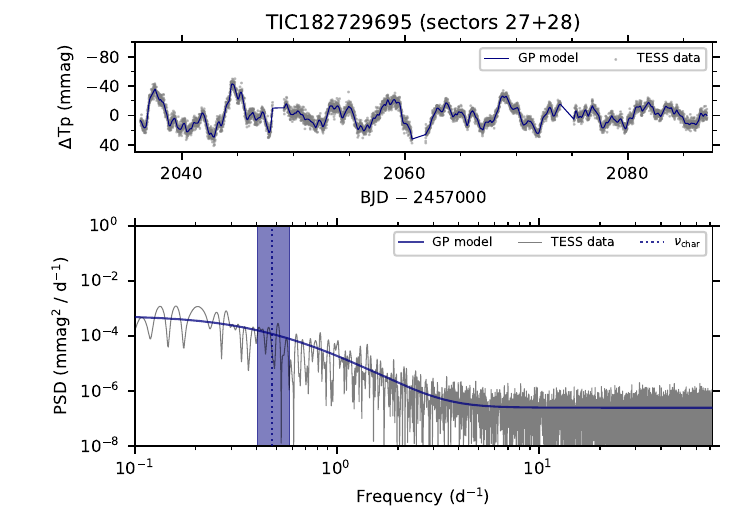}
\includegraphics[width=0.41\textwidth]{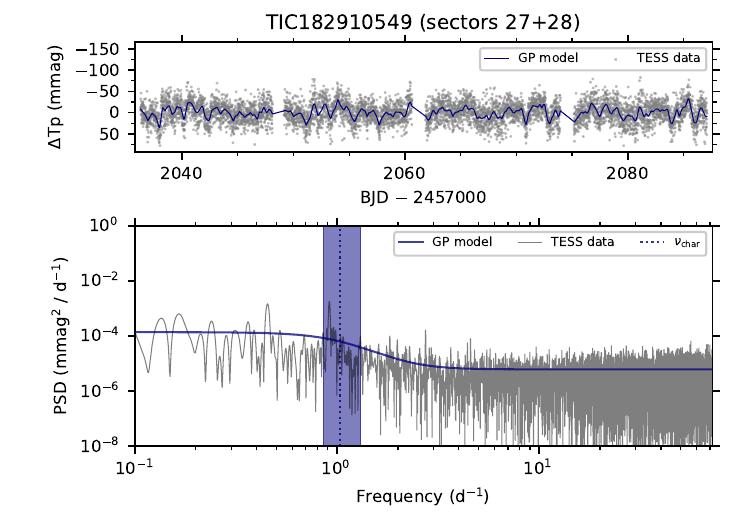}
\includegraphics[width=0.41\textwidth]{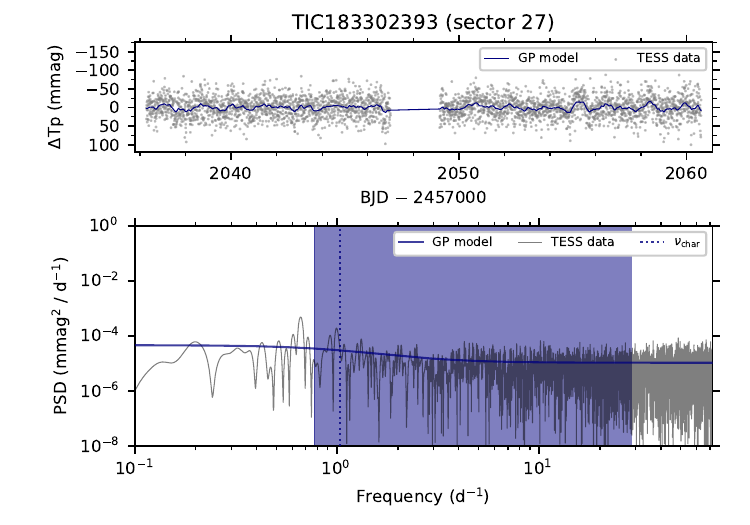}
\includegraphics[width=0.41\textwidth]{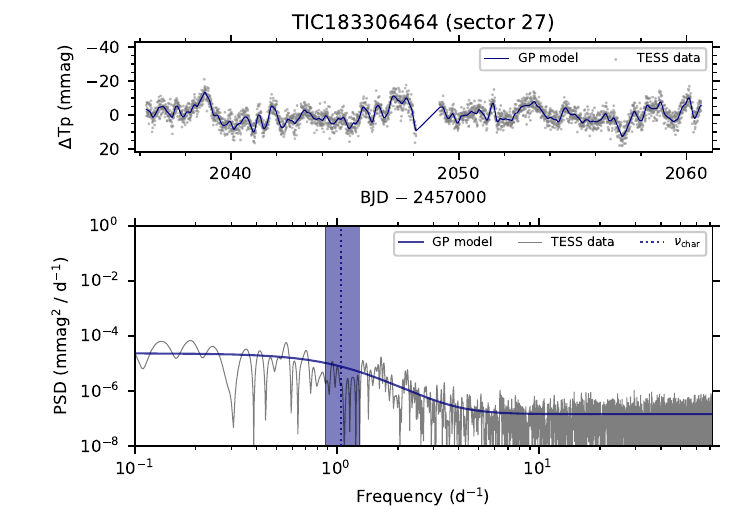}
\includegraphics[width=0.41\textwidth]{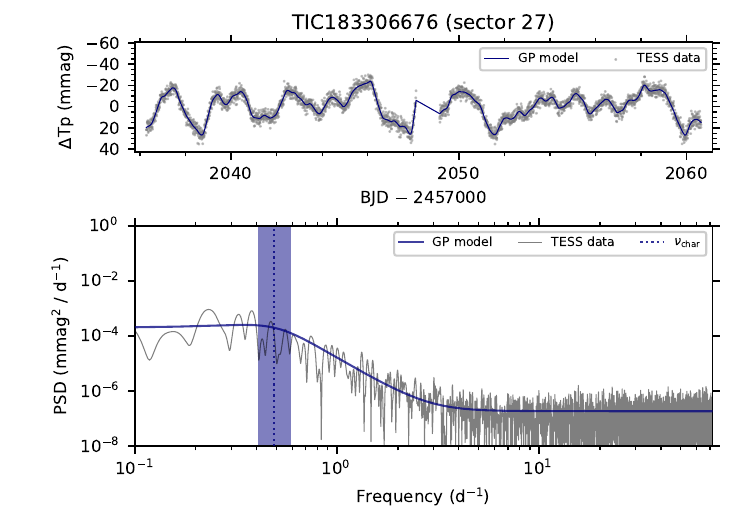}
\caption{Extracted ePSF light curves using {\tt tglc} \citep{Han_T_2023a} and fitted with the GP regression methodology of \citet{Bowman2022b} for SMC massive stars {\it (continued)}.}
\label{figure: SMC 2}
\end{figure*}

\begin{figure*}
\centering
\includegraphics[width=0.41\textwidth]{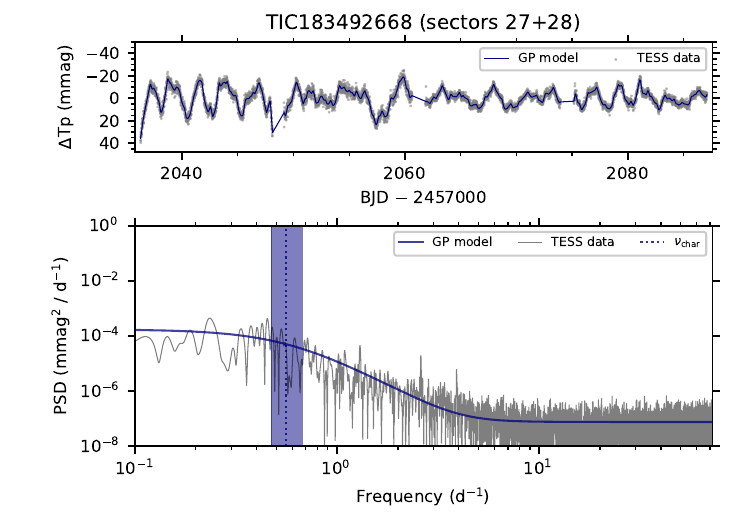}
\includegraphics[width=0.41\textwidth]{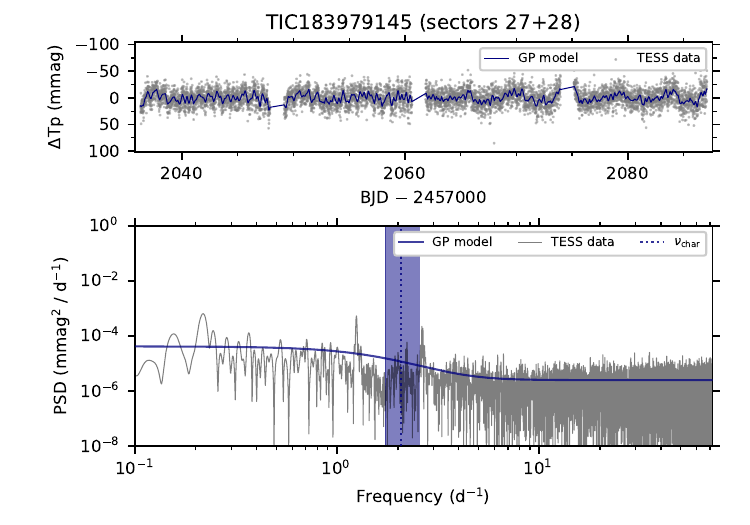}
\includegraphics[width=0.41\textwidth]{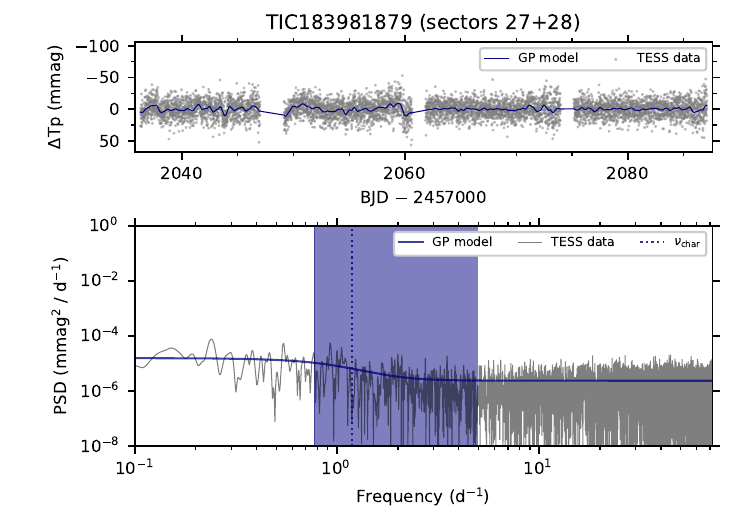}
\includegraphics[width=0.41\textwidth]{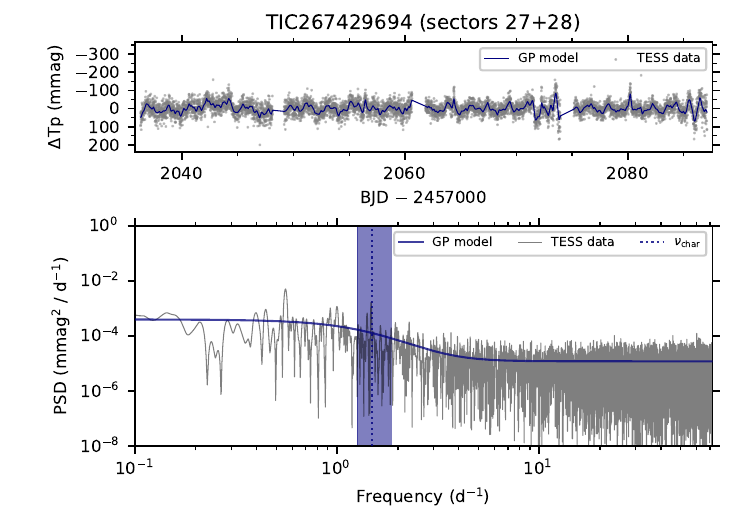}
\includegraphics[width=0.41\textwidth]{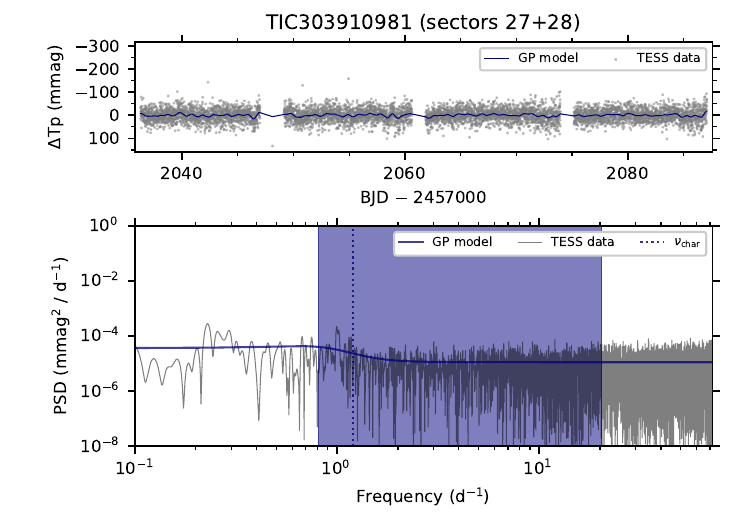}
\includegraphics[width=0.41\textwidth]{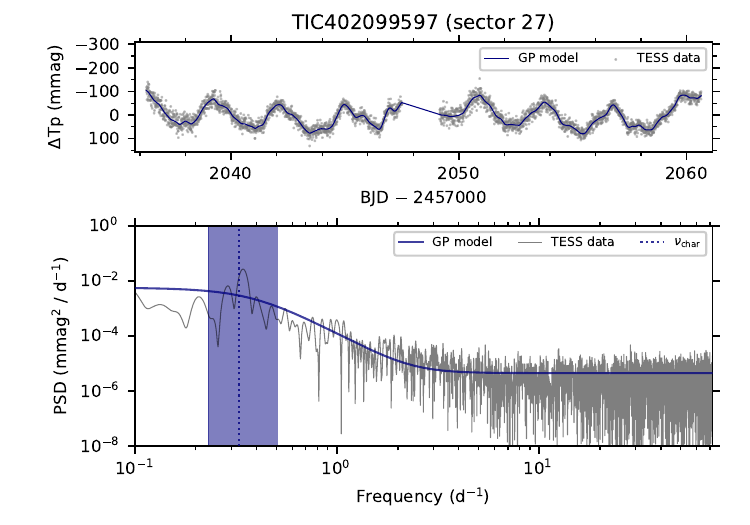}
\includegraphics[width=0.41\textwidth]{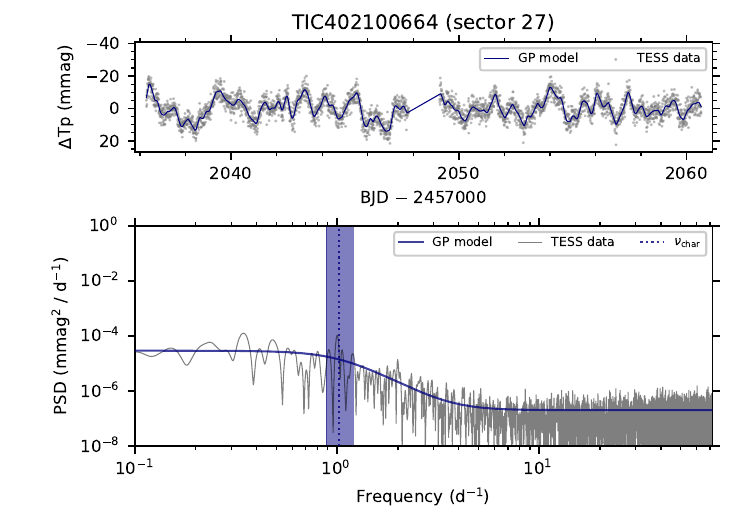}
\caption{Extracted ePSF light curves using {\tt tglc} \citep{Han_T_2023a} and fitted with the GP regression methodology of \citet{Bowman2022b} for SMC massive stars {\it (continued)}.}
\label{figure: SMC 3}
\end{figure*}

\begin{figure*}
\centering
\includegraphics[width=0.41\textwidth]{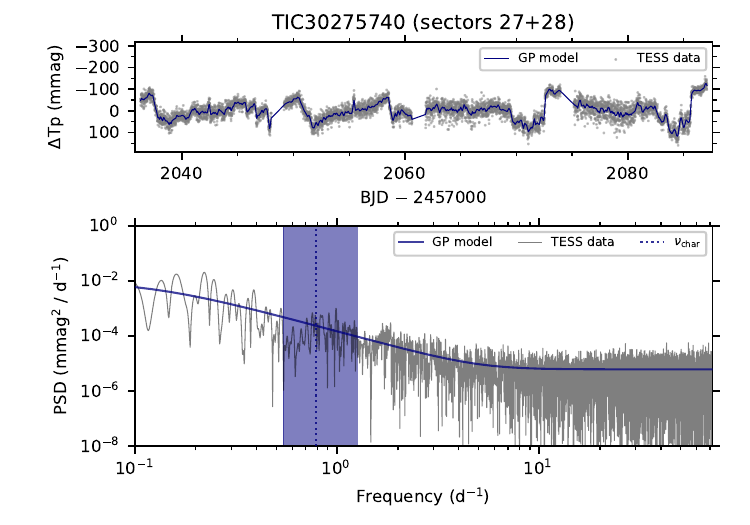}
\includegraphics[width=0.41\textwidth]{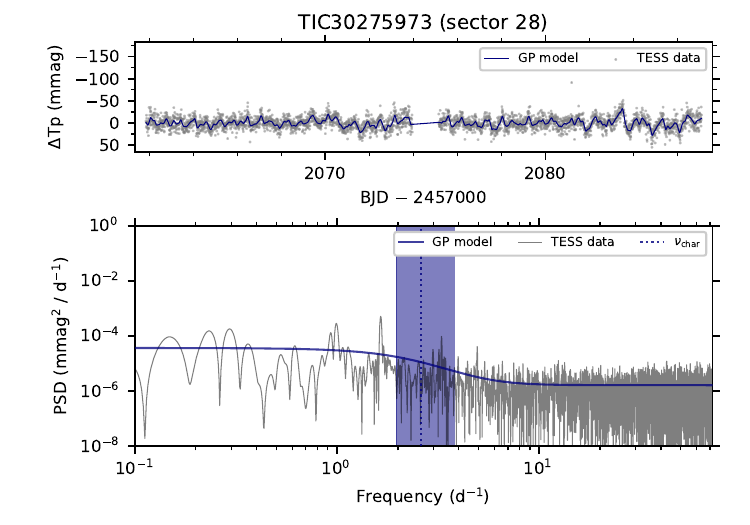}
\includegraphics[width=0.41\textwidth]{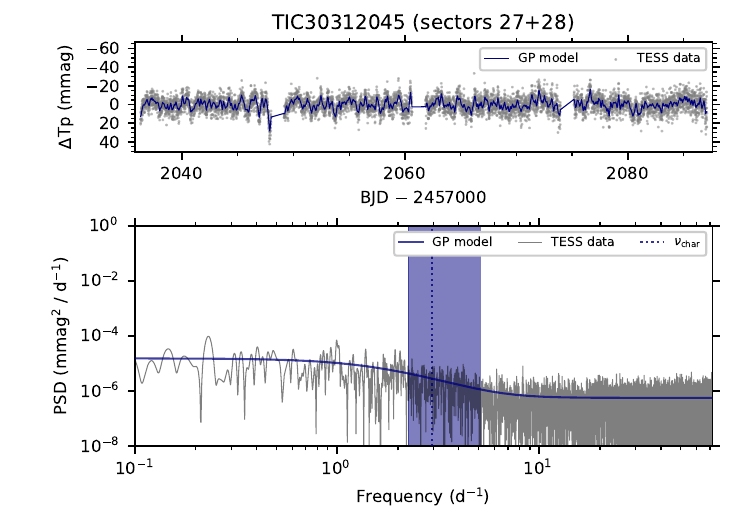}
\includegraphics[width=0.41\textwidth]{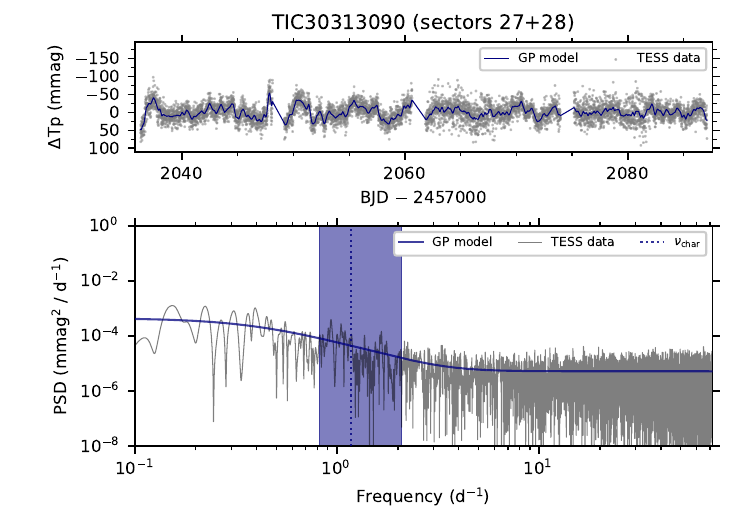}
\includegraphics[width=0.41\textwidth]{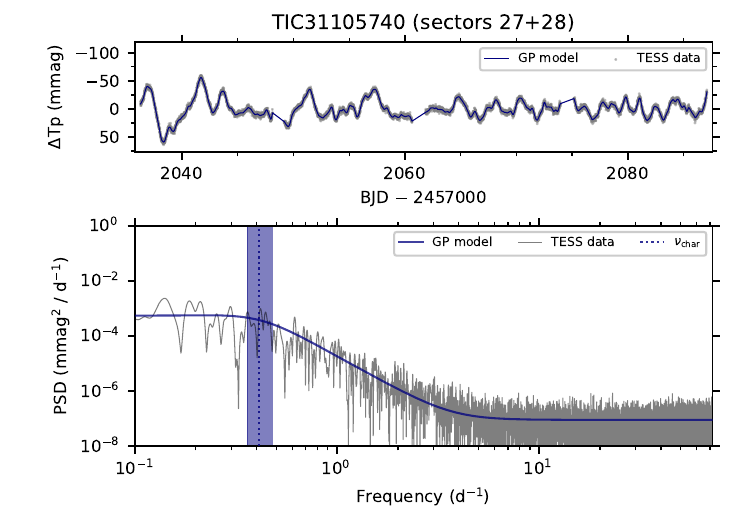}
\includegraphics[width=0.41\textwidth]{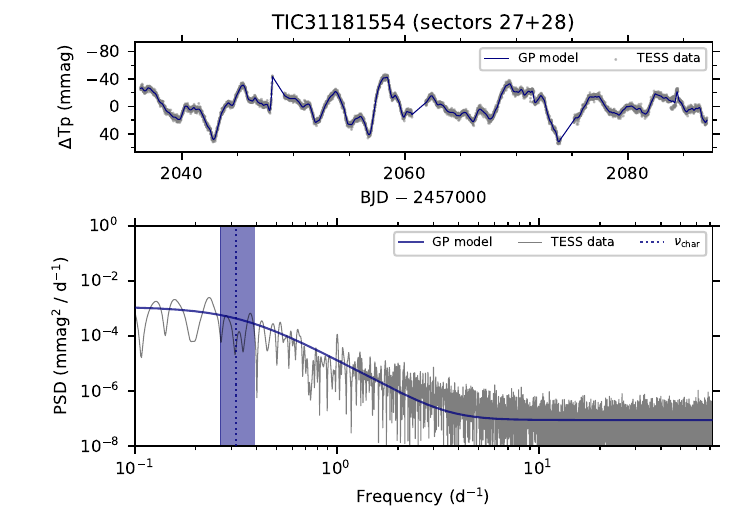}
\includegraphics[width=0.41\textwidth]{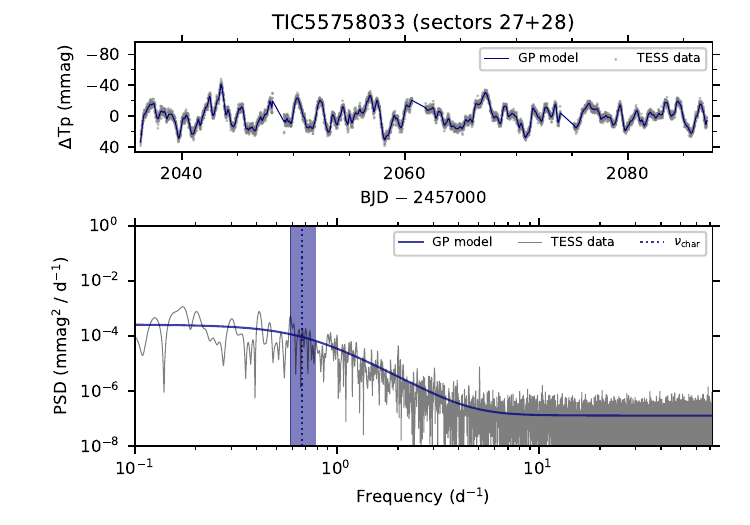}
\includegraphics[width=0.41\textwidth]{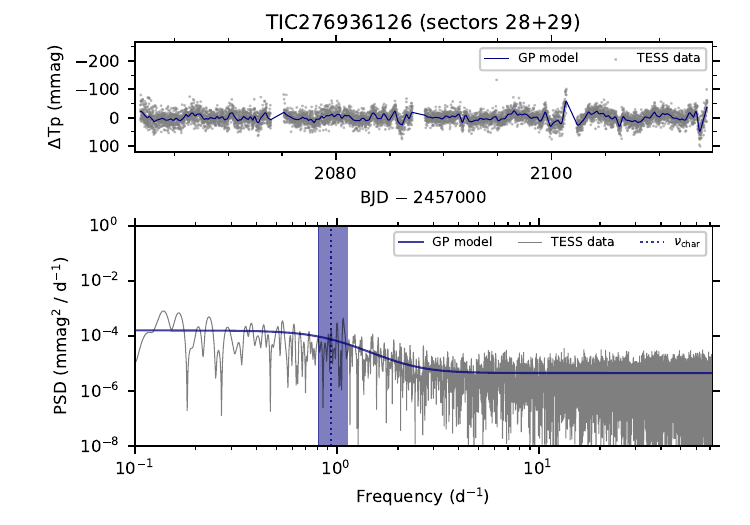}
\caption{Extracted ePSF light curves using {\tt tglc} \citep{Han_T_2023a} and fitted with the GP regression methodology of \citet{Bowman2022b} for LMC massive stars.}
\label{figure: LMC 1}
\end{figure*}

\begin{figure*}
\centering
\includegraphics[width=0.41\textwidth]{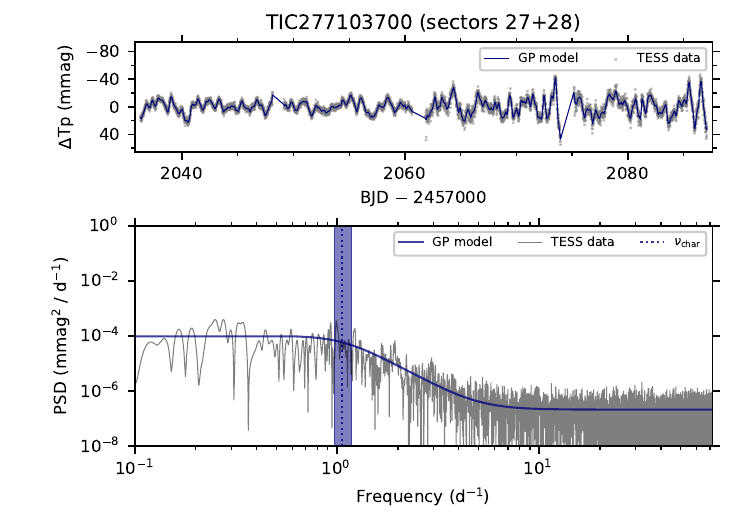}
\includegraphics[width=0.41\textwidth]{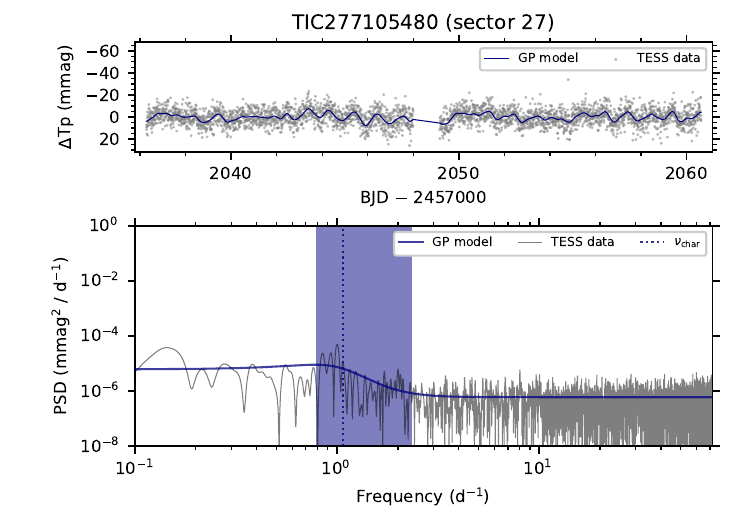}
\includegraphics[width=0.41\textwidth]{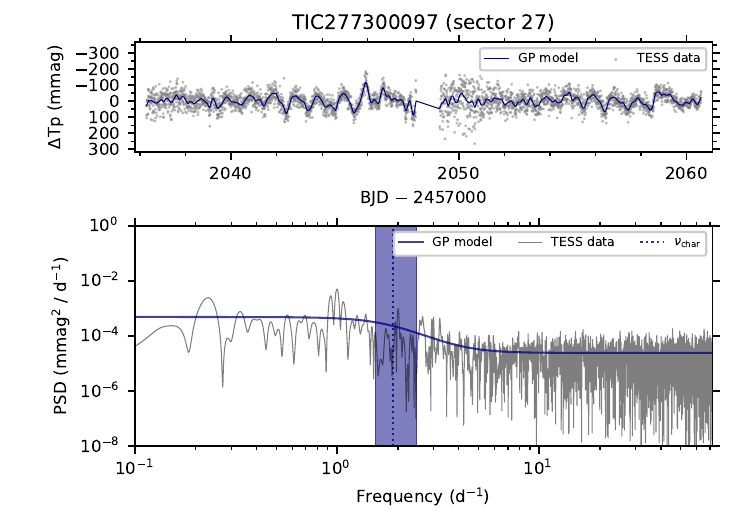}
\includegraphics[width=0.41\textwidth]{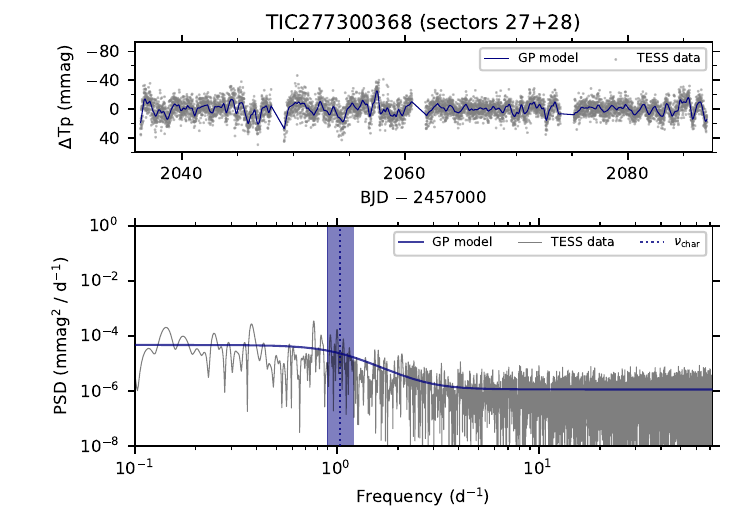}
\includegraphics[width=0.41\textwidth]{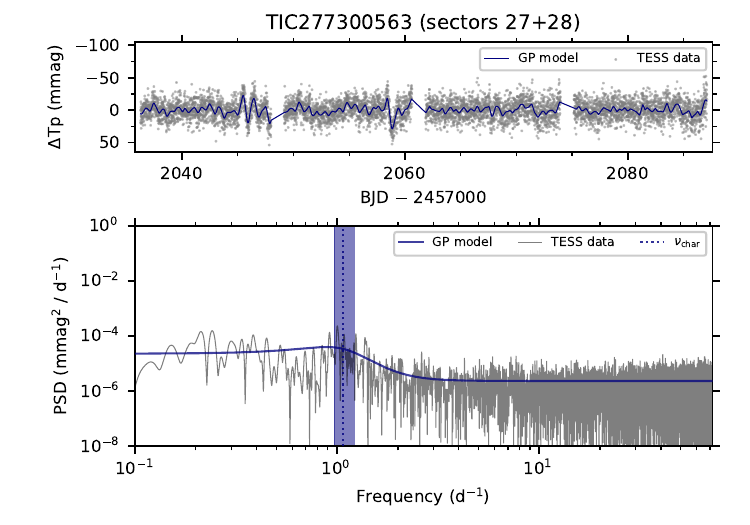}
\includegraphics[width=0.41\textwidth]{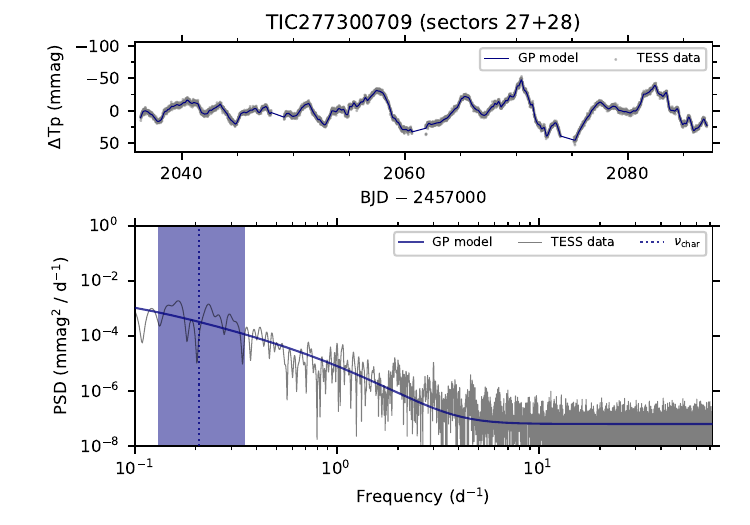}
\includegraphics[width=0.41\textwidth]{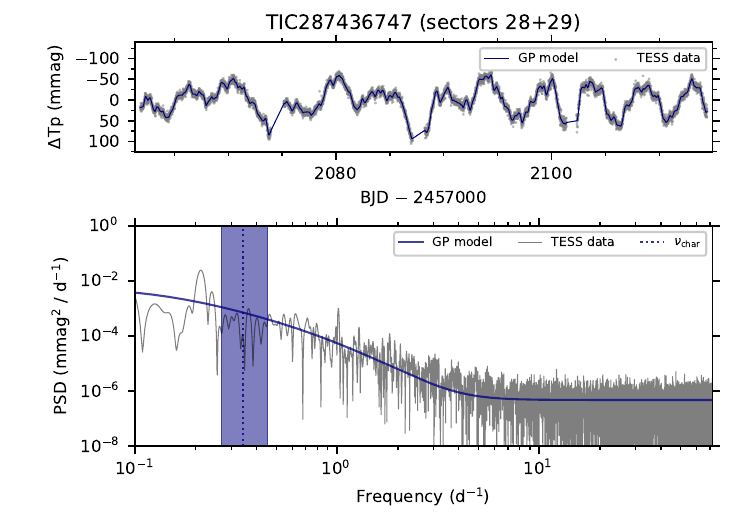}
\includegraphics[width=0.41\textwidth]{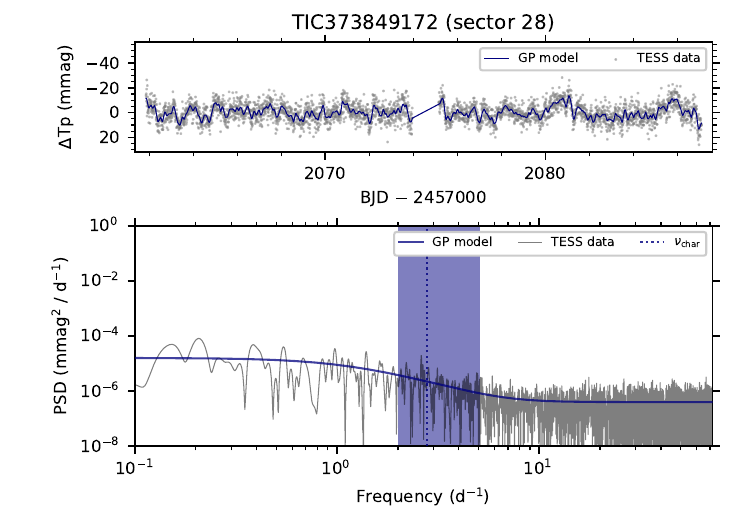}
\caption{Extracted ePSF light curves using {\tt tglc} \citep{Han_T_2023a} and fitted with the GP regression methodology of \citet{Bowman2022b} for LMC massive stars {\it (continued)}.}
\label{figure: LMC 2}
\end{figure*}

\begin{figure*}
\centering
\includegraphics[width=0.41\textwidth]{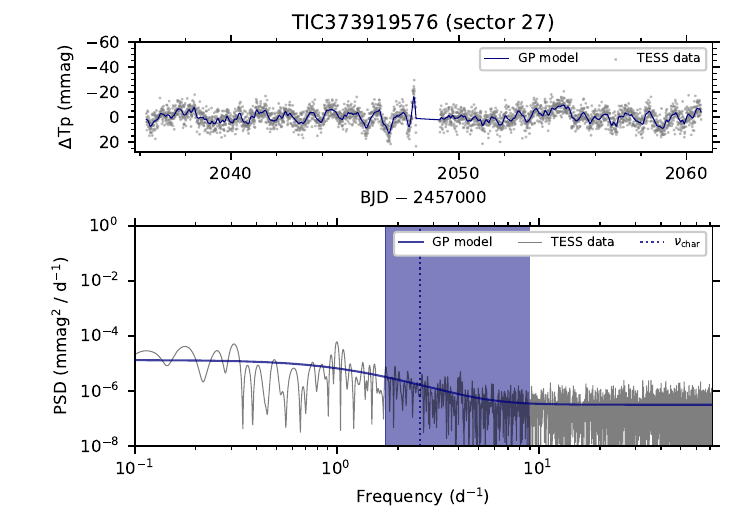}
\includegraphics[width=0.41\textwidth]{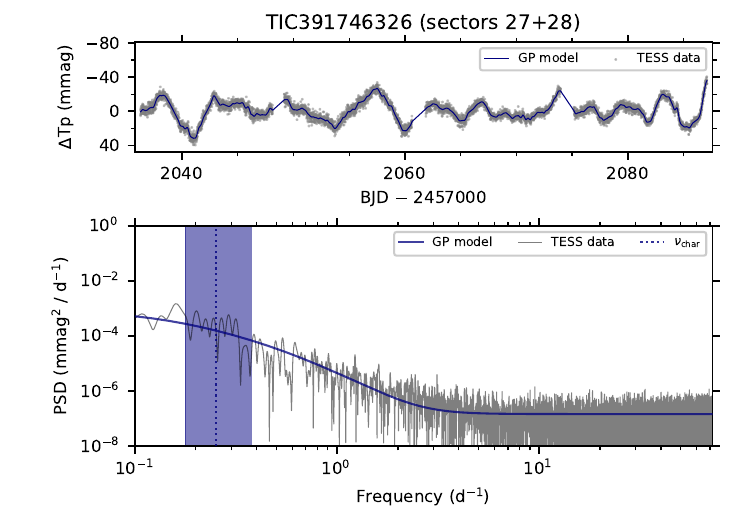}
\includegraphics[width=0.41\textwidth]{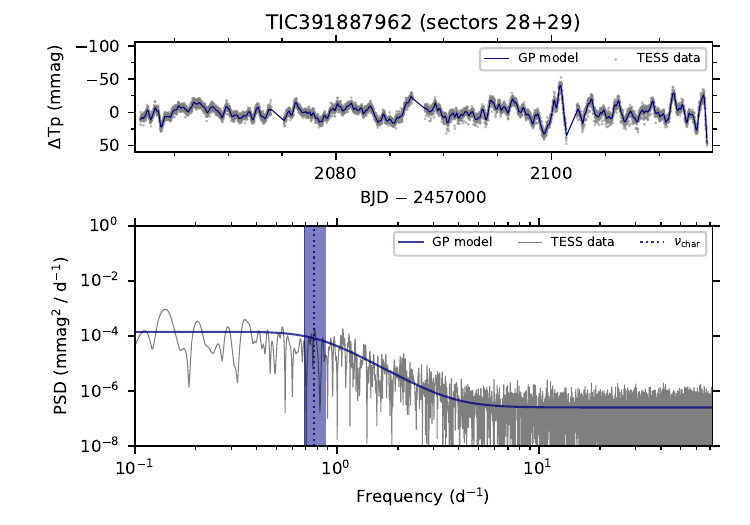}
\includegraphics[width=0.41\textwidth]{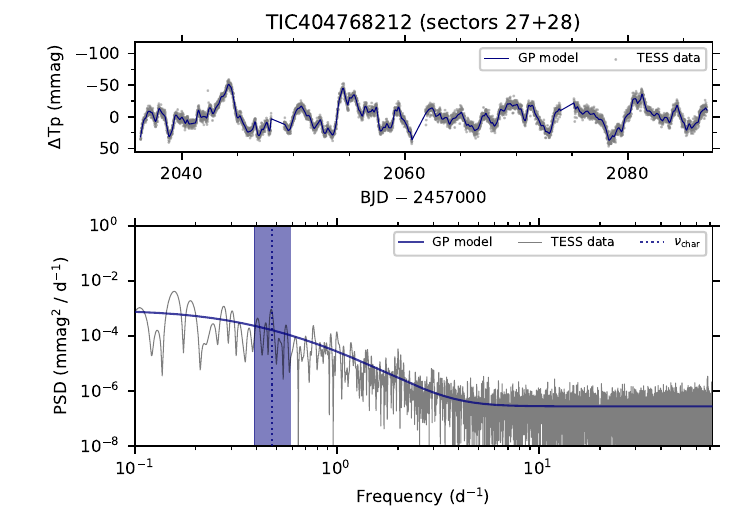}
\includegraphics[width=0.41\textwidth]{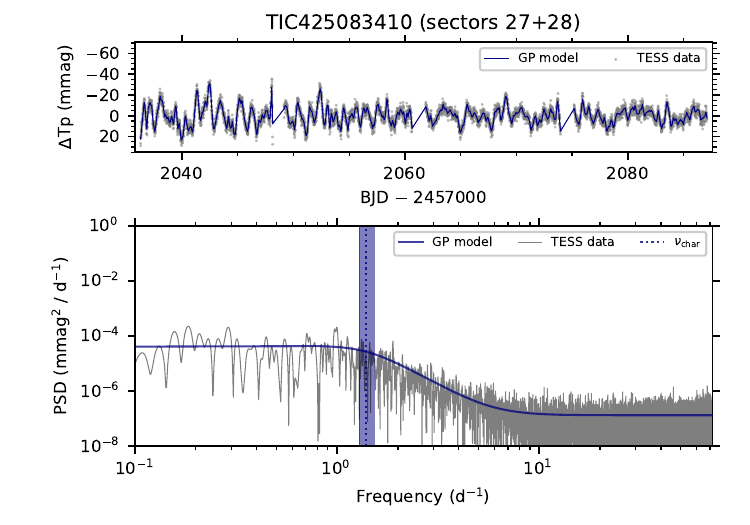}
\caption{Extracted ePSF light curves using {\tt tglc} \citep{Han_T_2023a} and fitted with the GP regression methodology of \citet{Bowman2022b} for LMC massive stars {\it (continued)}.}
\label{figure: LMC 3}
\end{figure*}

The extracted ePSF light curves using the {\tt tglc} software package \citep{Han_T_2023a} fitted with the GP regression methodology of \citet{Bowman2022b} for SMC and LMC massive stars are shown in Figs.~\ref{figure: SMC 1}--\ref{figure: SMC 3} and Figs.~\ref{figure: LMC 1}--\ref{figure: LMC 3}, respectively.

\end{appendix}


\end{document}